\documentclass[twocolumn]{aastex631}

\usepackage{amsmath}
\newcommand*\Bell{\boldsymbol\ell}

\newcommand\un[1]{{\,\rm #1}}
\newcommand\E[1]{\times10^{#1}}
\newcommand\rs[1]{_\mathrm{#1}}
\newcommand\g{$\gamma$}

\usepackage[dvipsnames]{xcolor}

\begin{document}

\title{On the properties of turbulence in the remnant of Tycho supernova}

\author[0000-0003-3487-0349]{Oleh Petruk}
\affiliation{INAF-Osservatorio Astronomico di Palermo, Piazza del Parlamento 1, 90134, Palermo, Italy}
\affiliation{Institute for Applied Problems in Mechanics and Mathematics, NAS of Ukraine, Naukova Street 3-b, 79060 Lviv, Ukraine}

\author[0000-0001-8464-0360]{Taras Kuzyo}
\affiliation{Institute for Applied Problems in Mechanics and Mathematics, Naukova Street 3-b, 79060 Lviv, Ukraine}

\begin{abstract}
We investigate the turbulent structures in the Tycho supernova remnant (SNR) by applying two-point autocorrelation and power spectral analyses to high-resolution X-ray and radio images. Using cleaned Chandra and VLA data, we derive two-dimensional (2D) and one-dimensional (1D) autocorrelation functions and power spectra, revealing that the fluctuations in both X-ray brightness (approximately tracing plasma density) and radio brightness (tracing magnetic field strength) follow the power-law distributions consistent with Kolmogorov-type turbulence. Specifically, power spectra exhibit a slope $8/3$ in 2D and $5/3$ in 1D analyses, indicating fully developed turbulence across a broad range of spatial scales. The structure functions and spectral slopes remain robust with and without correction for line-of-sight integration effects, suggesting the presence of Kolmogorov-like turbulence locally throughout the remnant's volume. Anisotropies in the 2D autocorrelation function are aligned with known density and magnetic field gradients. Analysis of the 1D radial properties of turbulence confirms Kolmogorov scaling and indicates that modes in the power spectra may be related either to periodicity in the fluctuations or to the physical sizes of bright or dim features in the surface brightness. Our results demonstrate that the Tycho SNR hosts a spatially and temporally developed turbulent cascade, providing insight into the physics of turbulence in SNRs.
\end{abstract}


\section{Introduction} 
\label{turb:intro}

The evolution of a star in the last epoch before it explodes as a supernova can be quite unstable, involving mixing between layers and eruption events. The structure of the circumstellar medium created by the stellar wind is also inhomogeneous. The supernova explosion creates a strong forward shock and turbulent ejecta, which expands with a supersonic speed and is perturbed by the reverse shock. All of these factors determine complex structures in spatial maps of supernova remnants (SNRs), as observed by various experiments. The quality of observational data in present-day or future astrophysical experiments from radio (LOFAR, SKA) through infrared and optical bands (HST, JWST) to X-rays (Chandra) allows one to see very detailed structures in the images of extended objects. In fact, the resolution of the instruments already goes down to the sub-arcsecond level, while supernova remnants could reach sizes of several degrees. With such instruments, we can study turbulence in SNRs on scales covering several orders of magnitude. 

The turbulent component of highly energetic plasma in SNRs is significant. It is dynamically important in the early stages, influencing their morphology, magneto-hydrodynamic and ionization structure, and the development of instabilities across various length scales. Consequently, it impacts cosmic ray injection, acceleration, and emission. Key open questions include the injection scale, the evolution and characteristics of the turbulence, the shape and progression of its spectrum, and similarities and differences among different SNRs.

\begin{figure*}
  \centering 
  \includegraphics[width=\textwidth]{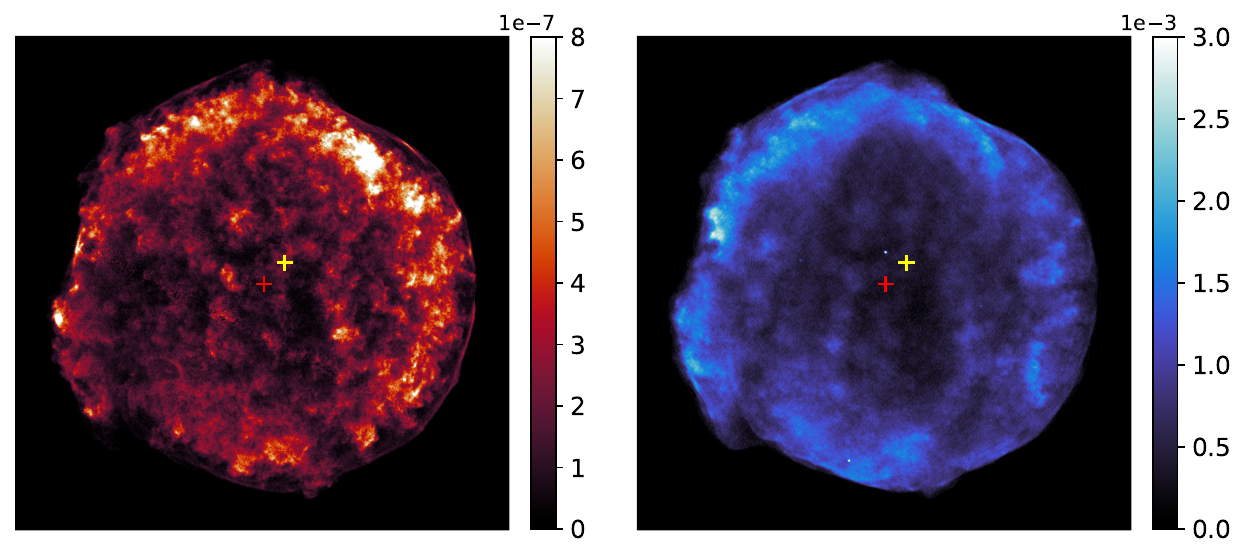}
  \caption{
   \textit{Left:} X-ray image of Tycho SNR in $1.2-4.0\un{keV}$, the surface brightness in units $\un{ph/(cm^{2}s)}$. \textit{Right:} The SNR's radio image at 1.4 GHz, in Jy/beam. The image sizes are $1220\times1220$ pixels ($10'\times 10'$). The images shown are after background noise removal as described in the Appendix~\ref{turb-app-fon}. 
   Crosses correspond to the geometrical center of the SNR \citep[the red one,][]{2005ApJ...634..376W,2010ApJ...709.1387K} and the actual explosion location \citep[the yellow one,][]{2015ApJ...809..183X}.%
  }
  \label{turb:fig-images}
\end{figure*}

Complex chaotic structures are seen in three-dimensional (3D) simulations of supernovae \citep[e.g.][]{2014ApJ...785..123C,2015A&A...577A..48W,2019MNRAS.485.3153B,2020ApJ...888..111O,2021MNRAS.503.4942O,2021MNRAS.502.3264G,2025ApJ...982....9V} and supernova remnants \citep[e.g.][]{2019A&A...622A..73O,2021ApJ...906...93F,2025arXiv250314455O}. Their morphology and pattern depend on different parameters involved in the simulations. To give some examples, we refer to the figure 3 in \citet{2012ApJ...749..156O} and to the figure 10 in \citet{2017ApJ...842...28W}. There are clear distinctions in the morphology of a random component in evolved remnants that have arisen from different initial radial profiles and ejecta clumpiness.

The analysis of properties of spatially distributed 'randomness' on different scales is widely used in cosmology by deciphering the power spectrum of cosmic microwave background radiation \citep[e.g.][]{2003ApJS..148..135H}. Similar methods are applied to structures that emerge in 3D simulations of SNRs  \citep{2021ApJ...906...93F,2022ApJ...940L..28P,2023ApJ...956..130M,2024ApJ...972...87M}. 
The authors use the angular power spectrum from decomposition to spherical harmonics in two dimensions (2D) or Legendre polynomials in one dimension (1D). The multipole expansion is also applied to the analysis of 3D numerical models of supernovae \citep{2020MNRAS.496.2039S,2021MNRAS.502.3264G}. 
The power spectrum from multiples characterizes the fluctuations pattern (anisotropy power) as a function of the angular scale. It is applied to spherical objects like CMB or a 3D SNR projection onto a sphere as seen from the explosion center. 

When studying the randomness in Galactic SNRs from their observed images (which are much smaller than the entire $4\pi$ celestial sphere), a slightly different approach is needed. Their images appear nearly flat, and one can analyze the power spectrum, which describes the amplitude of fluctuations as a function of the length scale. This spectrum can be derived from autocorrelation analysis.

A similar approach was recently used by \citet{prete2025} in a numerical study of the interaction between SNR and a turbulent ambient environment. 
In the present paper, we aim to decipher some properties of turbulence in the remnant of Tycho supernova by applying autocorrelation and power spectrum analysis to its observed X-ray and radio images. 

\begin{figure*}
  \centering 
  \includegraphics[width=\textwidth]{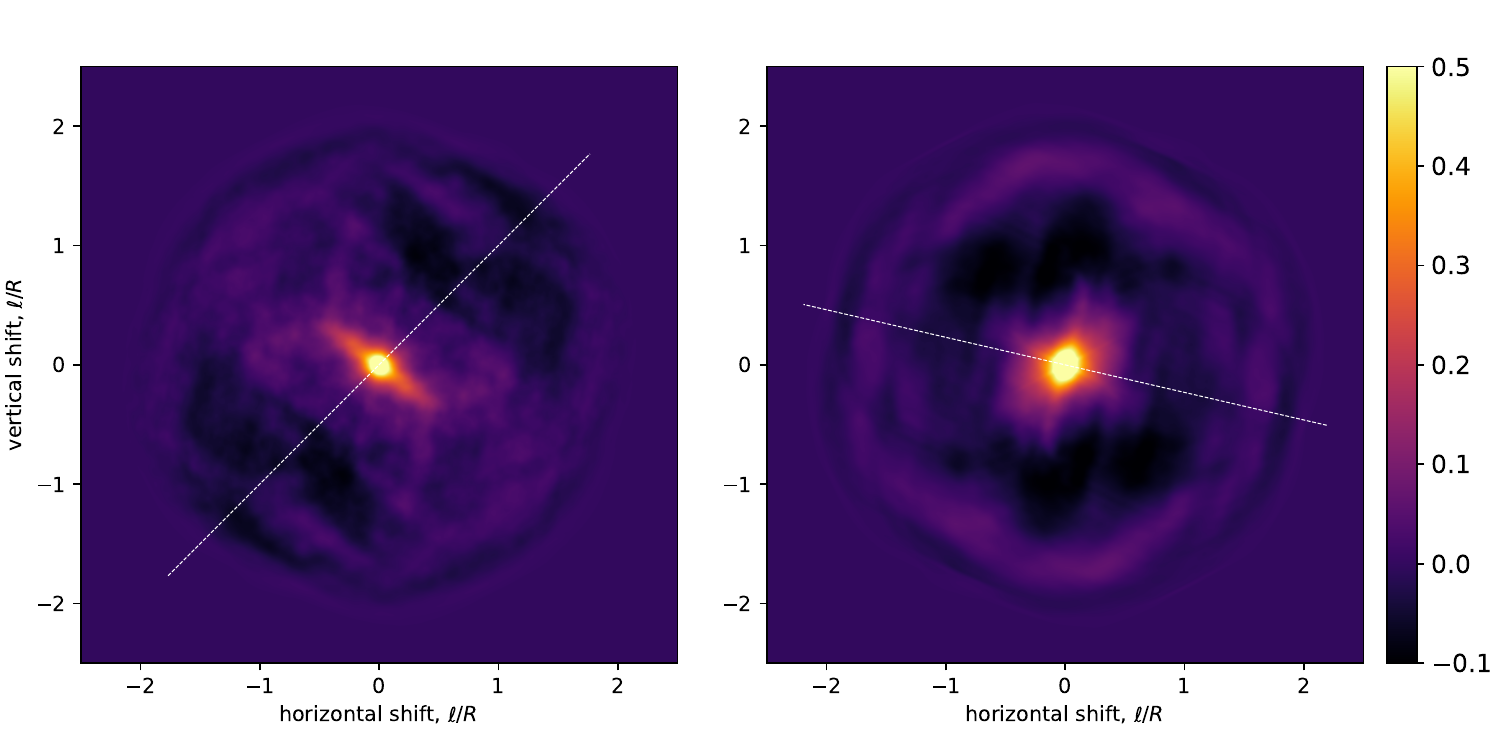}
  \caption{%
   2D autocorrelation functions for the X-ray image of Tycho SNR (\textit{left}) and for the radio image (\textit{right}) with the background noise removed. Numbers on axes are in units of the average radius of SNR $R=4'$. While lines correspond to directions of density (left) and magnetic field (right) gradients found by \citet{2024ApJ...972...63P}.%
  }
  \label{turb:fig-2Dautocor}
\end{figure*}

\section{Images and 2D autocorrelations} 
\label{turb:sect1}

\subsection{Derivation}

The present paper is based on observational data of Tycho SNR in X-ray and radio wavelengths. We analyze the X-ray image in photons with energies $1.2-4.0\un{keV}$. 
The image is derived from \textit{Chandra} data taken in 2015, obsid 15998, 147 ks of exposure \citep{2024ApJ...972...63P}. The pixel size in the image is $0.492''$. 
The radio image at 1.4 GHz was derived from VLA observations performed in
2013–2014 with the resolution $1.91''$ \citep{2016ApJ...823L..32W}. The pixel size is close to that in the X-ray image, $0.4''$. 
In order to allow for a direct comparison with the results from the X-ray observations, we have re-gridded the radio map to the X-ray map's resolution,  $0.492''$.

In the initial step, we have applied a standard noise-canceling algorithm to remove fluctuations that are not imminent to the SNR (background emission and instrumental noise). The procedure we followed is described in the Appendix~\ref{turb-app-fon}. In short, we calculated the 2D discrete Fourier transform (by using the Fast Fourier Transform algorithm) from the observed image and the background region of the same size ($10'\times10'$) in the SNR's vicinity. Then, we subtracted the background noise in the Fourier space
and performed the inverse Fourier transform to get the cleaned-up image of the SNR. 
The images after such a procedure are displayed in Fig.~\ref{turb:fig-images}. The differences between the original and the 'cleaned-up' images are practically unnoticeable, even in the radio images, where the signal-to-noise ratio is smaller. 
However, they become substantial later when we deal with the spectral properties of the autocorrelation on small scales. 

In the next step, we construct a 2D two-point autocorrelation function.
To do this, we use the Pearson definition of correlation, given by ${\mathrm{cov}(X, Y)}/{(\sigma_X \sigma_Y)}$ where $\textrm{cov}$ is a covariance, $X$ and $Y$ are random variables, and $\sigma_X$ and $\sigma_Y$ are their standard deviations. 
In our case, $X$ corresponds to a set of surface brightness values ${\cal I}$ from pixels located within the SNR boundary, while $Y$ refers to the same set linearly shifted in space, ${\cal I}'$.

\begin{figure*}
  \centering 
  \includegraphics[width=\textwidth]{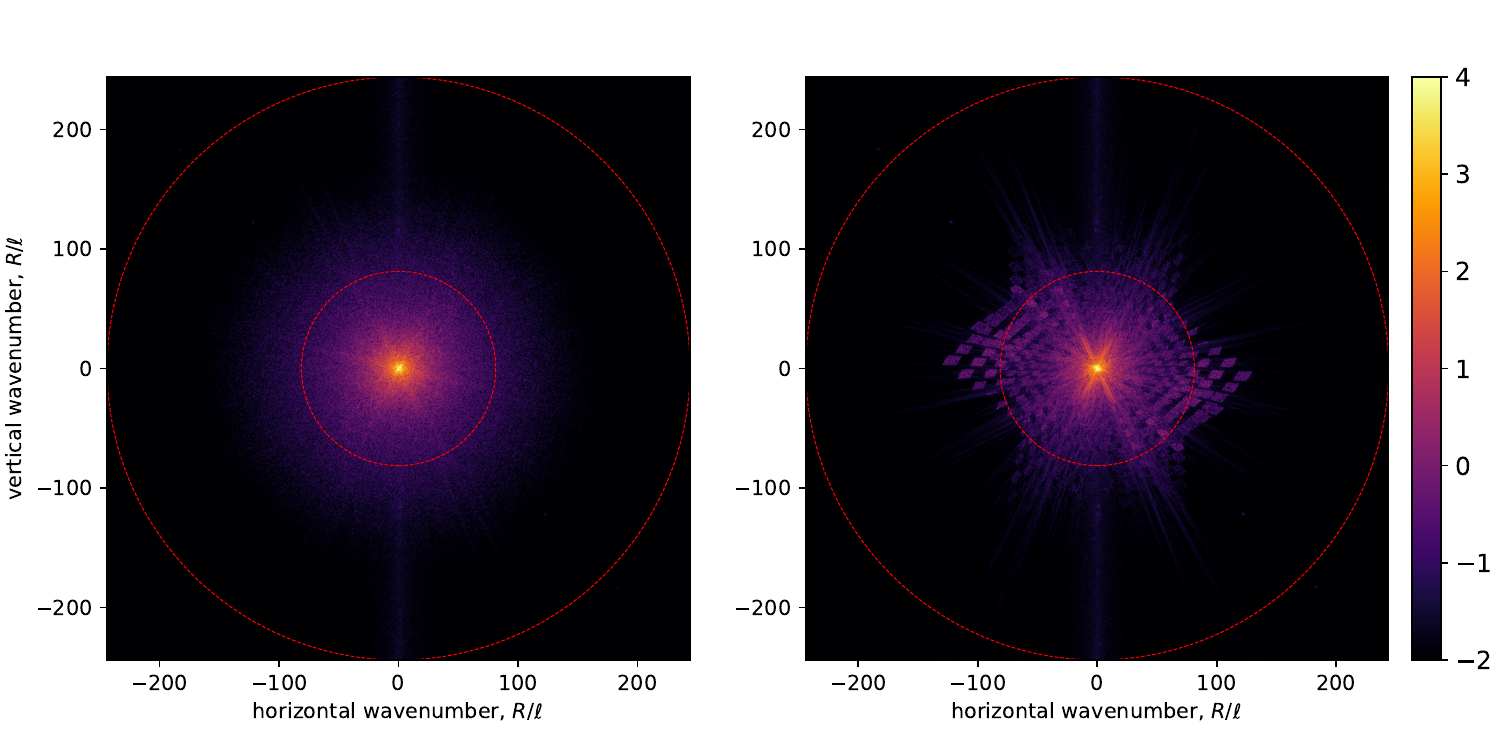}
  \caption{Discrete Fourier transform of the 2D autocorrelation maps from Fig.~\ref{turb:fig-2Dautocor}, for the X-ray (\textit{left}) and radio (\textit{right}) images. The color scale corresponds to the decimal logarithm of the Fourier amplitudes; the scale is the same for both images because they are derived from the autocorrelation functions, which are normalized to unity for the zero shift. 
   The inner red circle has a radius that corresponds to the wavenumber $k=1/3''$. The radius of the outer circle equals the Nyquist wavenumber $k=1/1''$.%
  }
  \label{turb:fig-2Dfourier}
\end{figure*}

\begin{figure}
  \centering 
  \includegraphics[width=0.95\columnwidth]{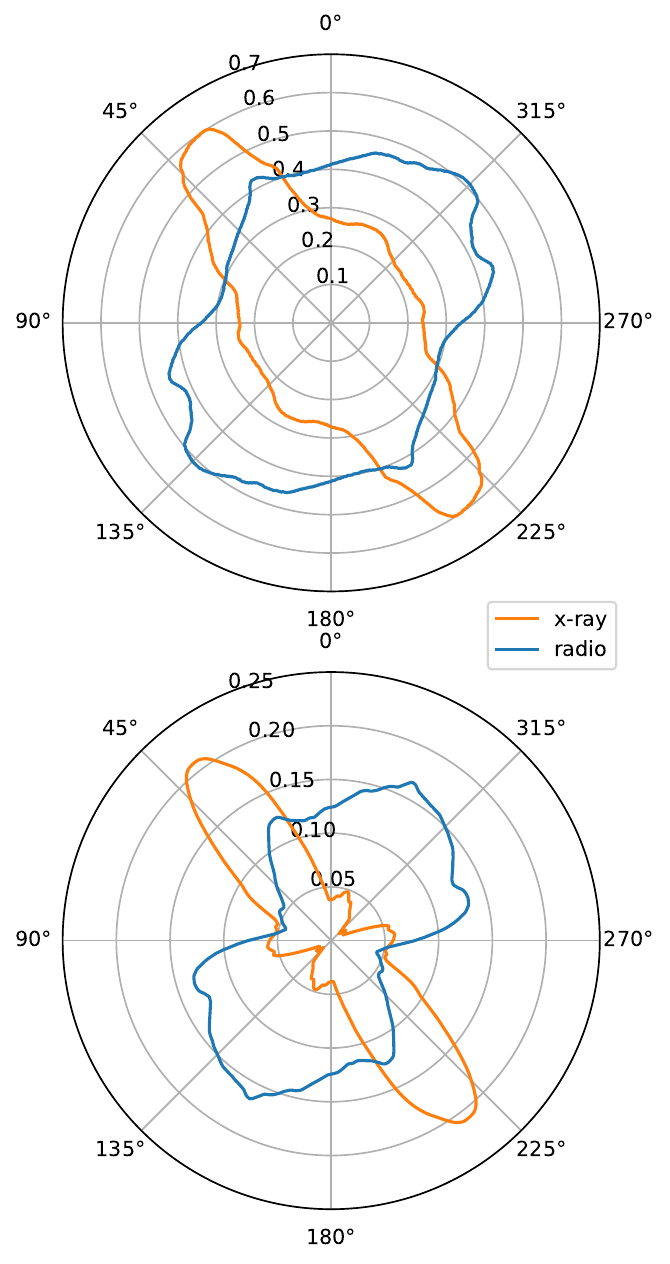}
  \caption{\textit{Top}: The correlation length $\ell_c$ (the distance where the correlation is $0.1$) for each direction as a fraction of $R$. \textit{Bottom}: the value of autocorrelation for a shift $1.5'$ ($0.375R$) as a function of the azimuthal direction.
   The orange lines correspond to the X-ray image and the blue lines to the radio image of Tycho SNR.%
  }
  \label{turb:fig-anizotr}
\end{figure}
\begin{figure}
  \centering 
  \includegraphics[width=\columnwidth]{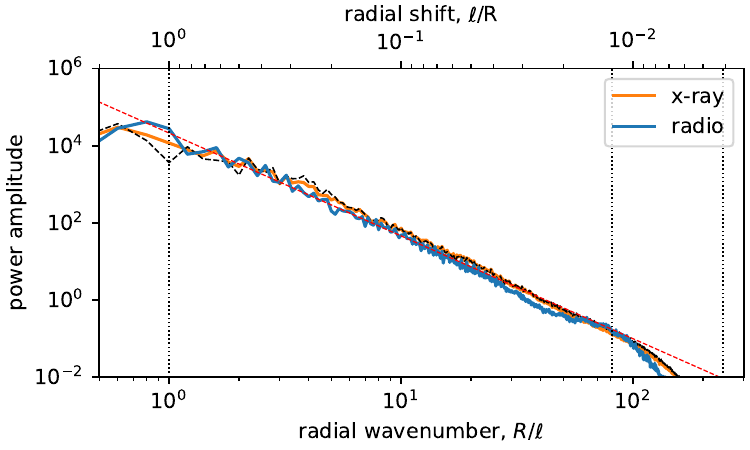}
  \includegraphics[width=\columnwidth]{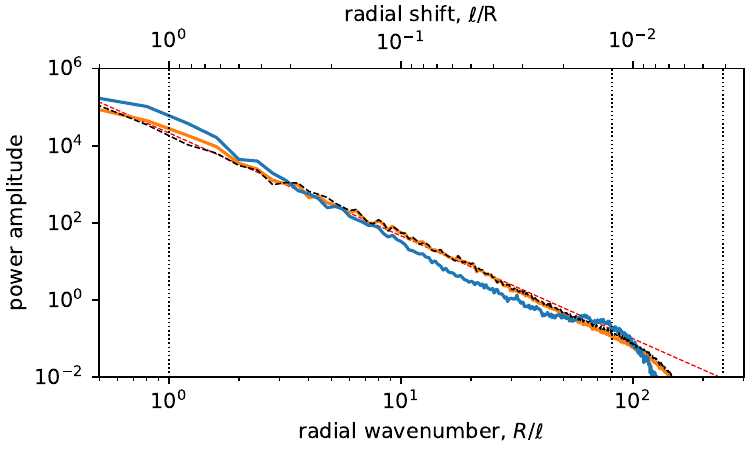}
  \caption{\textit{Top}: The radial power spectrum of the X-ray and radio brightness derived from FFT of the autocorrelation function. It is calculated from Fig.~\ref{turb:fig-2Dfourier} by averaging over the azimuth in the Fourier space. 
  \textit{Bottom}: The radial power spectrum calculated as the azimuth-averaged squares of the FFT amplitudes of the SNR images from Fig.~\ref{turb:fig-images}. 
  The amplitudes on the bottom plot are divided by $\sigma_{\cal I}^2 N$ to match the normalization of the top plot.
  Orange lines correspond to the X-ray image, and the blue lines correspond to the radio image of Tycho SNR. 
  Dashed black lines show the power spectrum for the X-ray image corrected for the geometrical factor, which reflects the integration along LoS. The red dashed line shows a power law with the index $8/3$ with the same normalization in both plots. 
  Vertical lines mark the wavenumbers corresponding to the average radius  $R=4'$ (left), a distance of $3''$ (middle), and the Nyquist wavenumber (right).%
  } 
  \label{turb:fig-1Dfourier}
\end{figure}

Namely, we calculate the map of the over-brightness $\delta {\cal I}={\cal I}-\overline{{\cal I}}$ where $\overline{{\cal I}}$ is the average brightness estimated from all pixels within the SNR's boundary. The boundary is determined from the radio map, which has a sharp outline marking the forward shock location. The values of $\delta {\cal I}$ outside this boundary are set to zero, so they do not contribute to the autocorrelation value. 
Then, we compute the sum of products of two over-brightness images while applying a relative shift between these images in all possible directions pixel-wise. As a result, we get a 2D autocorrelation as a function of the vector $\Bell$:  
\begin{equation}
 C_{\delta {\cal I}}(\Bell)
 =\frac{1}{\sigma_{{\cal I}}^2 N}\sum\limits_{(i,j)}^{N}\delta {\cal I}\cdot \delta {\cal I}'
 \equiv\langle\delta {\cal {\cal I}}\cdot \delta {\cal {\cal I}}'\rangle
\label{turb:defautocor} 
\end{equation}
where $\delta {\cal I}\equiv \delta {\cal I}(\mathbf{r})$, $\delta {\cal I}'\equiv \delta {\cal I}(\mathbf{r}+\Bell)$, the sum is over all pixels $\mathbf{r}=(i,j)$, $N$ is the total number of pixels, $\sigma_{\cal I}$ is the standard deviation for ${\cal I}$, $\mathbf{r}$ and $\Bell$ are 2D vectors. 
The maximum shift in a given direction corresponds to the SNR's diameter along this direction. 
The 2D autocorrelation function is centrally symmetric because the shifts by $\Bell$ and $-\Bell$ are equivalent (which follows from the ${\mathrm{cov}({\cal I}, {\cal I'})} = \mathrm{cov}({\cal I'}, {\cal I})$ symmetry).
Fig.~\ref{turb:fig-2Dautocor} shows the 2D two-point autocorrelation functions evaluated for the X-ray and radio images, respectively. The value in each pixel of this distribution represents the correlation coefficient between the original image of $\delta {\cal I}$ and the same image shifted to this point.

In the third step, we perform the Fourier transform of autocorrelation maps. 
According to the Wiener-Khinchin theorem, the result of the transform is the power spectrum of ${\cal I}$. 
The 2D plots for the power spectrum density of fluctuations in X-ray and radio images, derived as a 2D FFT of the 2D autocorrelation in Cartesian coordinates, are shown in Fig.~\ref{turb:fig-2Dfourier}. 
The large circles on this figure mark the Nyquist wave numbers (the highest wavenumber that could have a physical meaning). It corresponds to the size of the two pixels in the image, that is $2\cdot 0.492'' \approx 1''$.

Different sources of errors and how they could affect the results are discussed in Appendix~\ref{turb-app-errors}.

\subsection{Features}
\label{turb-sect22}

Fig.~\ref{turb:fig-anizotr} highlights the anisotropy in the 2D autocorrelation distributions from Fig.~\ref{turb:fig-2Dautocor} from two points of view. Namely, the plot on the top shows the correlation length as a function of the angular direction of the shift. We define the correlation length $\ell_c$ as the shift at which the correlation value is $0.1$. The bottom plot demonstrates the value of the correlation for a fixed spatial shift $\ell = 1.5'$ in all directions. 

There are a few properties apparent from these plots. 
We can see that in most directions, the correlation length is larger in the radio image. This means that the radio brightness changes less rapidly, and it shows smoother and less patchy features in these directions. 

The correlation is much stronger for shifts along the direction of approximately $36^\circ$ in the X-ray image and between $135^\circ$ and $150^\circ$ in the radio image. The most significant variations occur along the directions perpendicular to these, as indicated by the white dashed lines in Fig.~\ref{turb:fig-2Dautocor}. These directions correspond to gradients of density and magnetic field strength in the Tycho SNR \citep[cf. figure~1 in][]{2024ApJ...972...63P}. 
Indeed, in the considered range of the X-ray photon energies, the cooling function remains approximately constant, $\Lambda(T,\tau)\approx\mathrm{const}$ for temperatures $T\gtrsim 10^7\un{K}$ and a wide range of the ionization time-scales $\tau$ \citep[appendix in][]{2024ApJ...972...63P}. As a result, the X-ray image primarily reflects the properties of the density distribution $n$. In contrast, the radio emission is more sensitive to the magnetic field $B$, following ${\cal I}\propto nB^{1.6}$ if the radio spectral index is $0.6$, as measured in the Tycho SNR. The correlation distance from the image also has some traces of the density gradient, as one can see from the local maximum near $30^\circ$.

Fig.~\ref{turb:fig-1Dfourier} shows the power spectrum of disturbances in the X-ray and radio images of the SNR calculated in the two approaches. 
The top plot on Fig.~\ref{turb:fig-1Dfourier} is calculated from Fig.~\ref{turb:fig-2Dfourier}, i.e. from the Fourier transform of the 2D autocorrelation function, by averaging the 2D power spectrum over azimuth: $\overline{{\cal F}\left\{C_{\delta {\cal I}}(\ell)\right\}}$ where ${\cal F}$ marks a Fourier transform and the overline marks an average. 
There is another way to calculate the power spectrum, namely, by taking the azimuth-averaged squares of the Fourier amplitudes of the initial image from Fig.~\ref{turb:fig-images}, i.e., $\overline{{\cal F}\{\delta {\cal I}(r)\}^2}$. The power spectra obtained in this way are shown in the bottom plot of Fig.~\ref{turb:fig-1Dfourier}.
According to the Wiener–Khinchin theorem, these two approaches should yield the same power spectra. Indeed, the spectra on both plots are in perfect agreement. 

This Fig.~\ref{turb:fig-1Dfourier} demonstrates that the power spectra of fluctuations in the radio and the X-ray images follow the $k^{-8/3}$ law, which corresponds to the slope of the Kolmogorov turbulence in two dimensions. In particular, the spectra have nearly this shape up to the wavenumber $k\rs{*}\simeq 1/3''$ ($3''$ corresponds to 6 pixels). That is down to $\sim 1\%$ of the SNR radius (see upper horizontal axis), i.e., to the length scales involved in the acceleration of high-energy particles.
The spectra extracted from the SNR image decrease faster than power-law for $k>k\rs{*}$. The wavenumber $k\rs{*}$ marks a limit set by the instrumental possibilities rather than the beginning of the dissipation range. Indeed, the Reynolds number in SNRs $Re\simeq RV/\nu$ where $V$ is the shock speed and $\nu$ the magnetic viscosity is likely $Re\gtrsim 10^{8}$ \citep{1999ApJ...518..821R}, while the dissipation length scale $\ell\rs{dis}$ corresponds to $Re\simeq1$ with $Re\simeq \ell\rs{dis}v\rs{A}/\nu$. Therefore, roughly $\ell\rs{dis}\simeq \nu/v\rs{A}\lesssim 10^{-6}R$ for the same viscosity at different length scales and the Alfv\'en speed $v\rs{A}\simeq 0.01V$. 

What is remarkable in these results is that we derived the $k^{-8/3}$ spectrum for the entire SNR image. Brightness in different pixels is the sum of emissivities along the line of sight (LoS) inside the SNR. Generally, the turbulence spectrum can vary inside the actual remnant in a 3D space. Thus, the fact that we have derived the spectrum relevant for the 'fully developed turbulence' (i.e., the Kolmogorov one) means that the multiscale fluctuations in the ejecta of Tycho SNR are close to being statistically stable in time and space. In other words, the time scale for the cascade from the injection scale to the scale $\ell\rs{*}=1/k\rs{*}$ in Tycho SNR is much smaller than the age of the remnant. 

To perform an additional test, we calculated the geometric factor $\eta\rs{x}$ for each pixel over the X-ray image using the numerical hydrodynamic solution relevant for the remnant of an SNIa explosion at the age 450 yrs (see \citet{2024ApJ...972...63P} and Appendix~\ref{turb-app-geom-factor} for more details). Then we calculated the power spectrum from the X-ray image corrected for the geometrical factor. In this way, we eliminated the effect of integration along LoS. The derived spectra are shown on the top and bottom plots of Fig.~\ref{turb:fig-1Dfourier} with black dashed lines. 
In both approaches, the spectrum of the image corrected for $\eta\rs{x}$ has the same slope as the spectrum without this correction. This situation can occur when the local spectra along the line of sight have nearly the same slope throughout the SNR interior, and their sum naturally reproduces that slope. 

\begin{figure*}
  \centering 
  \includegraphics[width=0.92\textwidth]{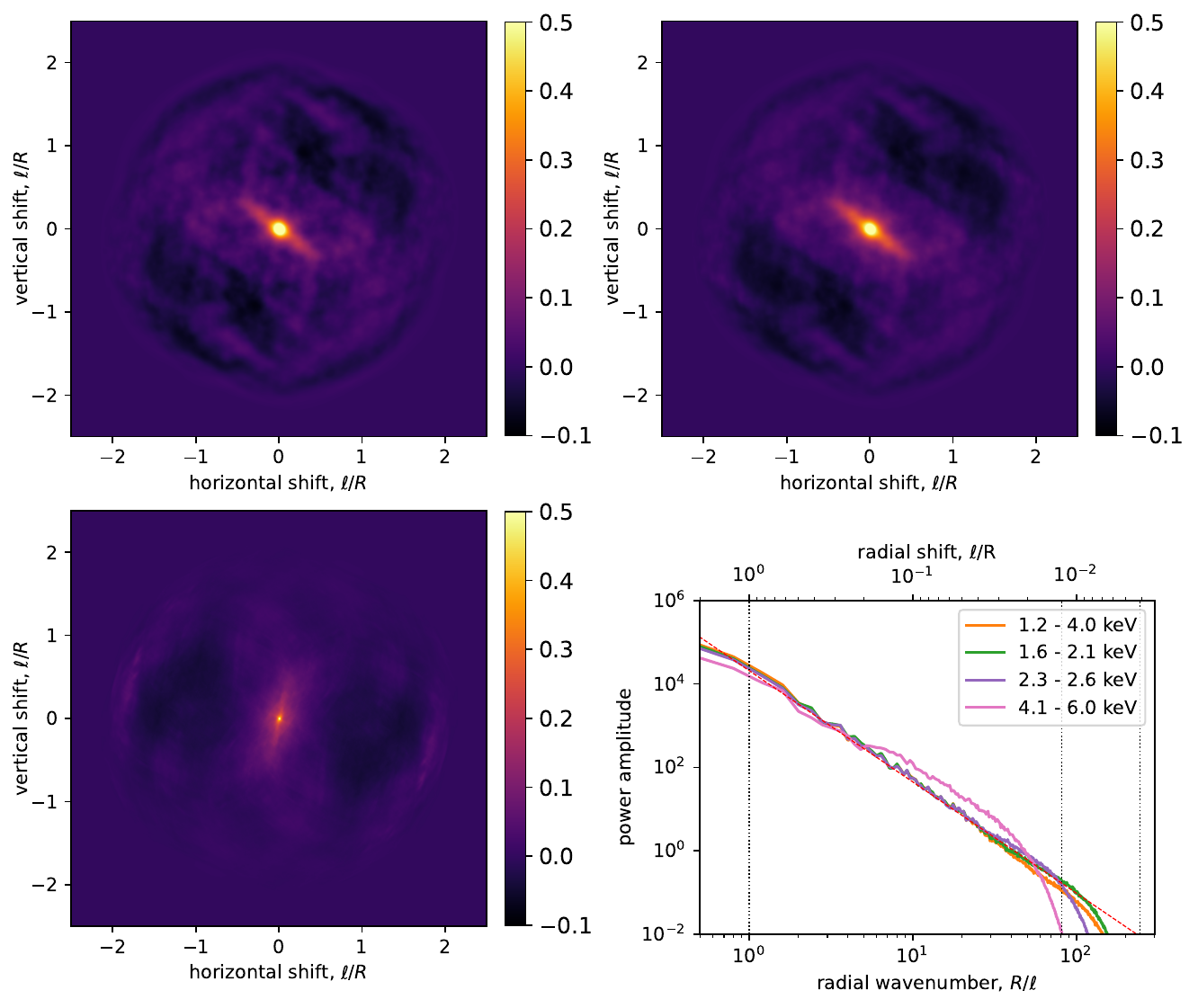}
  \caption{%
    2D autocorrelation functions for X-ray images of the Tycho SNR in 1.6-2.1 keV (upper left), 2.3-2.6 keV (upper right), and 4.1-6.0 keV energy range (bottom left).
    The bottom right plot shows the radial power spectra calculated as the angle average of ${\cal F}\{I\rs{x}\}^2$, for the SNR images $I\rs{x}$ in different photon energy ranges indicated in the legend. The dashed red line on this plot shows $k^{-8/3}$ dependence. The images were convolved before the analysis with kernels 1, 1, 1.5, 2.5 pixels as described in Appendix~\ref{turb-app-fon}.%
  }
  \label{turb:fig-xray-si-s}
\end{figure*}

\subsection{Narrow X-ray bands}
\label{turb-sect23}

In the previous subsections, the analysis is performed for the X-ray image in the photon energy range $1.2-4.0\un{keV}$. 
Let us repeat the same analysis for the images in the narrow bands where the X-ray emission from the two most prominent X-ray lines is dominant and in the hard X-ray continuum. Namely, we consider the images of Tycho SNR in photons with energies 1.6–2.1 keV (Si XIII), 2.3–2.6 keV (S XV), and 4.1-6.0 keV. The images were obtained from the same Chandra observations as above and have the same resolution and pixel size \citep{2025JPhSt..29.1901P}. 
Visually, the images in the two brightest lines are quite similar to the image in 1.2-4.0 keV, which is shown in Fig.~\ref{turb:fig-images}. The hard-band image differs \citep[e.g., figure~1 in][]{2011ApJ...728L..28E} because of the considerable contribution of the non-thermal emission. 
A new independent argument for this comes from the 2D autocorrelation functions shown in Fig.~\ref{turb:fig-xray-si-s}. The two upper plots on this figure are for the Si and S lines; they resemble very closely the 2D distribution for 1.2-4.0 keV (left plot in Fig.~\ref{turb:fig-2Dautocor}). In contrast, the direction of the highest correlation in the hard 2D autocorrelation function (left bottom plot in Fig.~\ref{turb:fig-xray-si-s}) is perpendicular to the MF gradient, as in the radio autocorrelation (right plot in Fig.~\ref{turb:fig-2Dautocor}).

The bottom right plot in Fig.~\ref{turb:fig-xray-si-s} shows the average 1D power spectra for the four X-ray ranges. 
These spectra follow the Kolmogorov law and are largely similar, except for the spectrum of fluctuations in the hard X-rays. Though it generally follows the $k^{-8/3}$ dependence, it demonstrates some deviations.

Considering the tight similarity between the thermal autocorrelations and spectra as well as the essential difference with hard X-rays, we may safely assume negligible contamination from the non-thermal emission in the range 1.2-4.0 keV, which we use in our analysis.

\begin{figure}
  \centering 
  \includegraphics[width=\columnwidth]{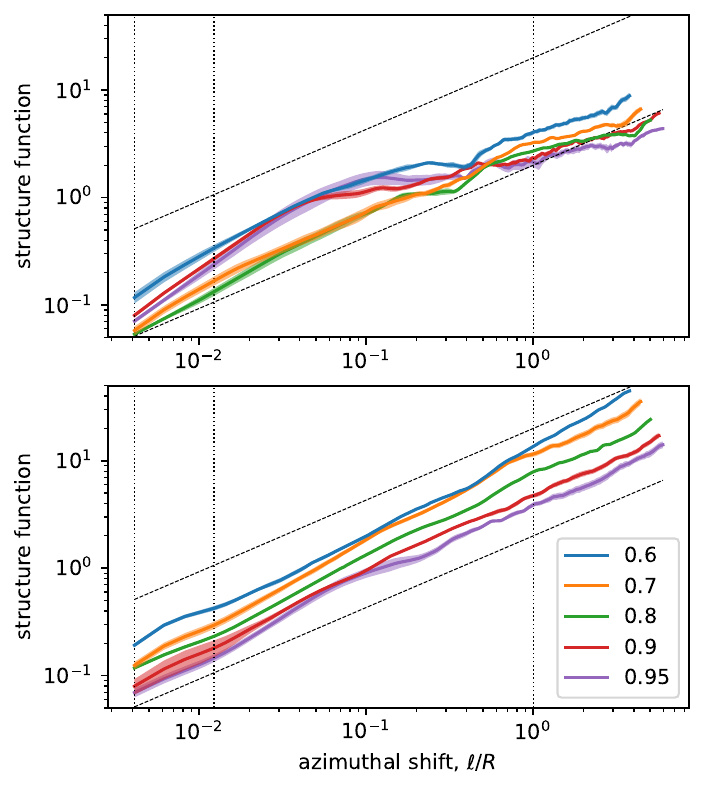}
  \caption{Function $S_{\delta n}(\ell)$ calculated for the X-ray (\textit{top}) and the radio (\textit{bottom}) images of Tycho SNR from a few contours consisting of pixels at the distances $r/{\cal R}$ (numbers in the legend) where ${\cal R}(\varphi)$ is the radius from the explosion center that varies with azimuth.
  The plots are approximately representative of the plasma density fluctuations (upper) and the magnetic disturbances (lower). The dashed lines show the power law $\ell^{2/3}$. 
  The vertical lines mark the wavenumbers $k=1/1''$ (left), $k=1/3''$ (middle), and corresponding to the average radius $R=4'$ (right). 
  The horizontal axis is in units of $R=4'$. The maximum shift along the contour corresponds to the total length of that contour.
  The $1\sigma$ statistical errors for the structure function are also shown (by lighter colors around solid lines).%
   }
   \label{turb:fig-stripes-spectrum}
\end{figure}

\section{Structures in density and magnetic field}
\label{turb:sect-spectrum}

\subsection{Structure function}

By analyzing the radio image of the Tycho SNR, \citet{2018MNRAS.480.2200S} have found that the magnetic disturbances in this object are of the Kolmogorov type. 
We follow their approach to relate correlations in the X-ray surface brightness to the ejecta density fluctuations. 
Namely, we analyze the thermal X-ray image of Tycho SNR in photons with energies $1.2-4.0\un{keV}$ where the cooling function $\Lambda(T,\tau)\approx\mathrm{const}$. The thermal X-ray surface brightness ${\cal I}\propto n^2\Lambda\eta$, where the filling factor $\eta$ accounts for the emissivity variation along the LoS. 

The length of LoS inside an SNR varies across the projection. To reduce the geometric effect on the correlation function, \citet{2018MNRAS.480.2200S} considered the brightness over the concentric {\em circles} from the {\em geometrical} center of the SNR, assuming that the length of LoS and $\eta$ are likely the same at the same radii. 

In this section, we correlate the values of the brightness ${\cal I}$ 
from contours with a shape that differs from a circle. Namely, we consider thin (thickness $0.02R$) stripes at the {\em relative} radii $0.7{\cal R}, 0.8{\cal R}, 0.9{\cal R}, 0.95{\cal R}$ where the radius of the SNR ${\cal R}(\varphi)$ is measured from the {\em explosion} center and varies with azimuth $\varphi$. In order to determine ${\cal R}$, we used the radio image, which yields a more precise location of the forward shock. 

We start from the autocorrelation function for the surface brightness ${\cal I}$ 
\begin{equation}
 C_{\cal I}(\ell)\equiv\langle {\cal I}(r)\cdot {\cal I}(r+\ell)\rangle.
\end{equation}
If ${\cal I}\propto n^2$ approximately holds, then we may derive an expression for the second-order structure function of the plasma density fluctuations (Appendix~\ref{turb-app-deltan-derivation}, equations (\ref{turn-app-eq3})-(\ref{turn-app-eq5}))
\begin{equation}
 S_{\delta n}(\ell)\equiv\langle(\delta n-\delta n')^2\rangle_n\approx\frac{C_{\cal I}(\ell\rs{min})-C_{\cal I}(\ell)}{c_1}
\label{turb:eq-structfunc} 
\end{equation}
where $\delta n=n-\bar n$, $\bar n$ is the average density,  
$c_1$ is a constant. 

\citet{2018MNRAS.480.2200S} have analyzed the radio image of the Tycho SNR.
Assuming no fluctuations in the density of emitting electrons, the authors have derived approximately $C_{\cal I}(\ell\rs{min})-C_{\cal I}(\ell)\propto \ell^{2/3}$ for the magnetic field disturbances across 1D shapes they considered (their fig.~2). The index $2/3$ corresponds to the turbulence considered by \citet{1941DoSSR..30..301K,1962JFM....13...82K}. 

We apply such analysis to the surface brightness taken from our contours and calculate $\overline{C_{\cal I}(\ell\rs{min})-C_{\cal I}(\ell)/c_1}$ where the overline marks the average over the thickness of the stripes. 
Fig.~\ref{turb:fig-stripes-spectrum} shows the dependence $S_{\delta n}(\ell)$ for several distances from the center, for the X-ray and radio images of Tycho SNR. 

\begin{figure}
  \centering 
  \includegraphics[width=\columnwidth]{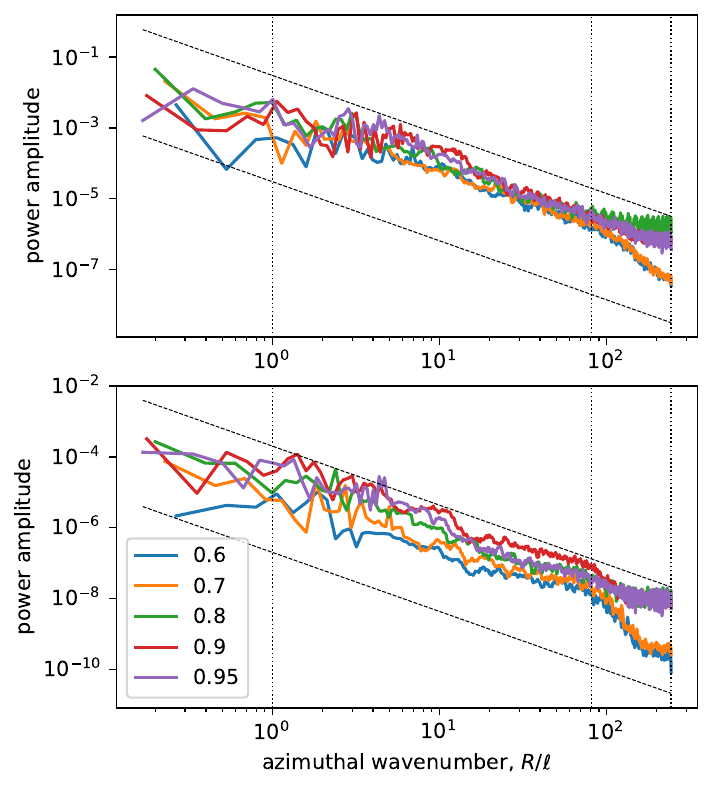}  
  \caption{Power spectrum nearly representative for the density ({\em top}) and magnetic field ({\em bottom}) fluctuations in Tycho SNR calculated as a square of the Fourier transform of ${\cal I}\rs{x}^{1/2}$ and ${\cal I}\rs{r}^{1/1.6}$ respectively across the same contours as used for Fig.~\ref{turb:fig-stripes-spectrum}. 
  To reduce the noise out of the main dependence, we have applied a Wiener filter \citep[][sect.~9.2.4]{1990ph...book.....L} with window size 9 pixels to the spectra.
  The dashed lines show the decreasing power law with index $5/3$. 
  The vertical lines mark the wavenumbers $1/R$ (left), $1/3''$ (middle), and $1/1''$ (right).%
   }
   \label{turb:fig-stripes-spectrum-fourier}
\end{figure}

For the X-ray image, we found that the density fluctuations are close to the $S_{\delta n}(\ell)\propto \ell^{2/3}$ dependence. The structure function deviates somehow from this law for $r=0.9{\cal R}$ and $0.95{\cal R}$. The reason could be that the ejecta (which dominates the X-ray emission in the considered photon energy range) is sampled less accurately at these distances, and the spectrum is probably modified by the instabilities at the contact discontinuity, as suggested by \citet{2018MNRAS.480.2200S}. 

For the radio image, we recovered the results of \citet{2018MNRAS.480.2200S} although we considered contours with different shapes. 
In Sect.~\ref{turb-sect22}, we saw that the radial variation of $\eta$ (i.e., the geometric effect due to integration along LoS) does not considerably modify the power spectrum in Tycho SNR. The similarity of the structure functions derived by \citet{2018MNRAS.480.2200S} and in the present paper, from different contours, confirms this point. It seems that such a low sensitivity to the 3D geometry of an SNR could be natural if there is turbulence of the \citet{1941DoSSR..30..301K,1962JFM....13...82K} type, i.e., with random local values of MHD parameters.  

Even more, \citet{1941DoSSR..30..301K} dealt with a 3D system and derived the $\ell^{2/3}$ dependence for 1D shifts along a radial direction in polar coordinates. Therefore, generally speaking, one should prove that the same $\ell^{2/3}$ law is valid for a contour, i.e., for the azimuthal dependence. However, we can hypothesize that the dependence will be nearly the same for any smooth contour for the Kolmogorov-type turbulence.

\subsection{Power spectrum}

In his original papers, \citet{1941DoSSR..30..301K,1962JFM....13...82K} has demonstrated that the 1D structure function ${\cal S}$ is proportional to $\ell^{2/3}$ for the stationary, locally uniform, and isotropic turbulence without inflow or losses of energy. \citet{1949RSPSA.199..238B} have shown that the power spectrum for the Kolmogorov-like turbulence should be proportional to $k^{-5/3}$ in the 1D space. 

What are the power spectra of fluctuations for the same contours as we used above to calculate the structure function? Again, there are two equivalent approaches. It may be calculated either as a Fourier transform of the two-point autocorrelation function or as the square of the Fourier amplitudes.

In the first approach, we could calculate it from the SNR image as (Appendix~\ref{turb-app-deltan-derivation}, equation (\ref{turn-app-eq4})) 
\begin{equation}
 C_{\delta n}(\ell)\equiv\langle \delta n\cdot \delta n'\rangle\approx
 1-\frac{C_{\cal I}(\ell\rs{min})-C_{\cal I}(\ell)}{2c_1},
 \label{turbeq:eq-corr-n}
\end{equation}
that is, in fact $C_{\delta n}(\ell)=1-S_{\delta n}(\ell)/2$.  
If the function $S_{\delta n}\propto \ell^{2/3}$ then the Fourier transform for positive $k$ 
\begin{equation}
 {\cal F}\left[1-\frac{A\cdot(\ell^2)^{1/3}}{2}\right](k)=\
 \frac{A}{\sqrt{6\pi}}\,\Gamma\left(\frac{2}{3}\right)k^{-5/3},
\end{equation} 
where $A$ is a constant and $\Gamma(x)$ is the gamma function, provides the slope $5/3$ for the power spectrum of fluctuations in 1D. 

Instead, we calculate the power spectrum in the second approach, namely, from the direct Fourier transform. 
From the X-ray image, by taking ${\cal I}\rs{x}\propto n^2$, we compute the power spectrum for the density fluctuations as 
$\overline{{\cal F}\left\{{{\cal I}\rs{x}(r)^{1/2}}\right\}^2}$
where the overline marks the average over the thickness of a stripe. It is shown on the top plot in Fig.~\ref{turb:fig-stripes-spectrum-fourier}.  
The bottom plot in the same figure shows the power spectrum for the magnetic field obtained from the radio image as
$\overline{{\cal F}\left\{{{\cal I}\rs{r}(r)^{1/1.6}}\right\}^2}$
by assuming the synchrotron brightness ${\cal I}\rs{r}\propto B^{1.6}$. 
We see from Fig.~\ref{turb:fig-stripes-spectrum-fourier} that the shapes of spectra on both plots are close to $k^{5/3}$. 
Such similarity for different functions $\overline{{\cal F}\left\{{{\cal I}\rs{x}(r)^{1/2}}\right\}^2}$ and $\overline{{\cal F}\left\{{{\cal I}\rs{r}(r)^{1/1.6}}\right\}^2}$ is not of surprise because we deal with the Kolmogorov-type turbulence where the values of MHD parameters (and therefore of emissivity and brightness) are almost random locally (on the smallest scale). 

\begin{figure*}
  \centering
  \includegraphics[width=0.82\textwidth]{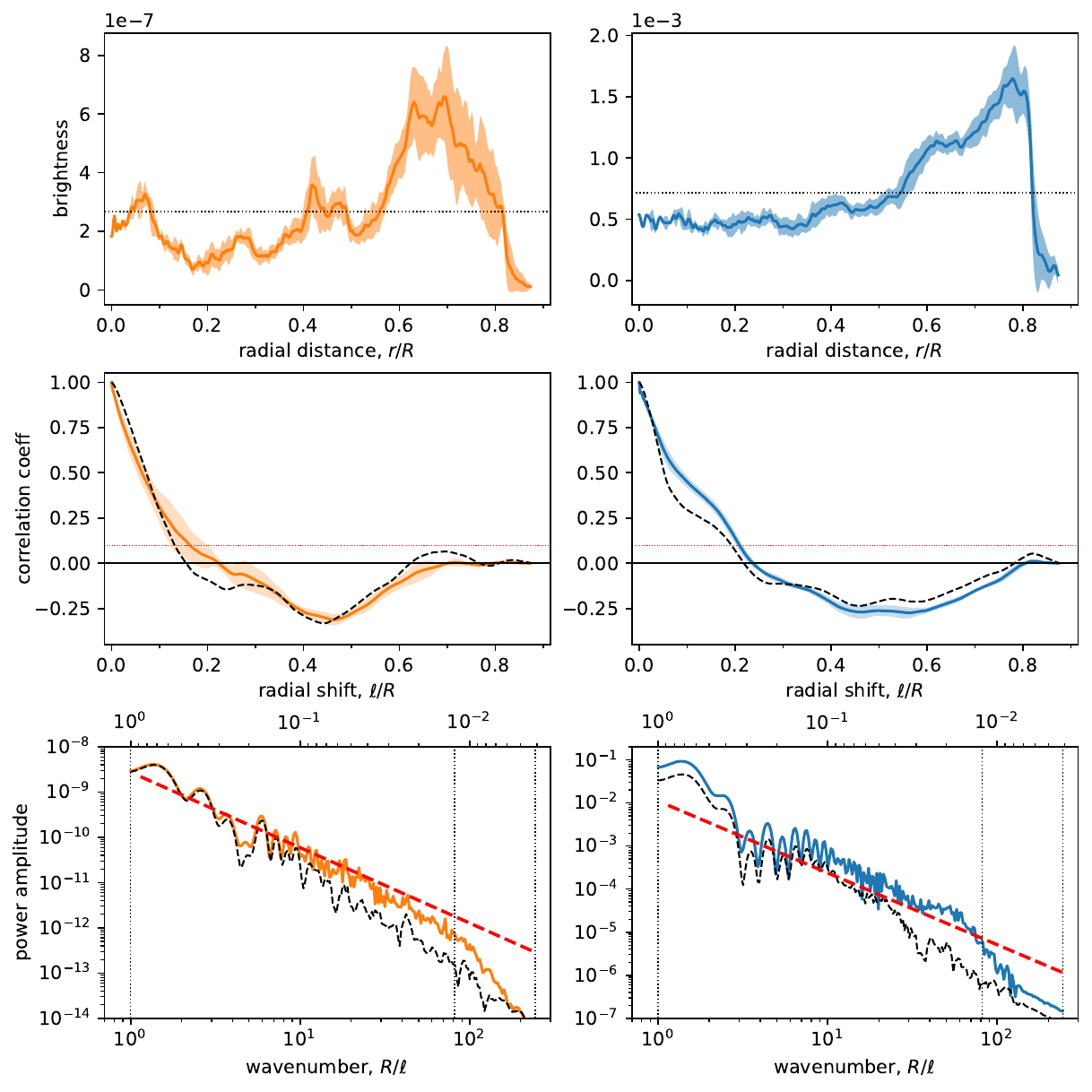}
  \caption{%
       Profile of the surface brightness ${\cal I}$ (top plot), autocorrelation function $C_{\delta {\cal I}}$ (middle), and the power spectrum ${\cal F}\left\{{{\cal I}(r)}\right\}^2(k)$ (bottom) for the sector $330^\circ$ - $345^\circ$ (measured counterclockwise from the North). Left column for the  
       X-ray and the right column for the radio images.
       Brightness is averaged over azimuth by taking 45 individual radial profiles within the sector. 
       The autocorrelation function and the power spectrum are calculated individually from each of these 45 brightness profiles and then averaged. The errors shown are at the $1\sigma$ level. The horizontal line on the plot for ${\cal I}$ shows the average brightness. The red horizontal line on the plot for $C_{\delta {\cal I}}$ corresponds to correlation 0.1. The red dashed line on the power spectrum is the power law with the index $5/3$. 
       Vertical lines on the power spectrum plot mark wavenumbers $1/R$, $1/3''$, and the Nyquist one. Spatial scales are normalized to the average radius $R = 4'$. 
       The black dashed lines in the middle row represent the autocorrelation function calculated from the average brightness profile.
       The same lines in the bottom plots are the power spectra from the average brightness profiles, with a Wiener filter (window size 4 pixels) applied to diminish the noise in the spectrum.%
   }
   \label{turb:fig-effect-wiener}
\end{figure*}

\section{1D autocorrelation functions}
\label{turb:sect-1Dauto}

\citet{1941DoSSR..30..301K} considered features of turbulence along the radial directions. Let us explore the radial properties of fluctuations in the Tycho SNR. More specifically, we divide the SNR image into several sectors and calculate the power spectra and the autocorrelation functions for the radial shifts along their radial directions.  
The maximum shift in this case is the radius of the SNR, not its diameter, as it is in Sect.~\ref{turb:sect1}. 

Fig.~\ref{turb:fig-effect-wiener} shows the results of the calculations for one sector. The products for a number of other sectors are presented in Appendix~\ref{turb-app-1D-radial-autocorr}.

We see from Figs.~\ref{turb:ap:fig-sectors-xray-fft} and \ref{turb:ap:fig-sectors-radio-fft} in this Appendix that the power spectra for the radial brightness profiles are close to a power-law with index $5/3$, confirming again that the turbulence in Tycho SNR is of the Kolmogorov type. The autocorrelation functions and spectra differ among sectors, which provides evidence that features in the disturbances vary with azimuth. 
We leave a detailed analysis of them for the future, but would like to highlight several features. 

There is a sort of cutoff in the power spectra around the same wavenumber, $1/3''$, as we have seen in Fig.~\ref{turb:fig-1Dfourier}. It has, seemingly, an instrumental origin. 

The spectra are generally closer to the power law at wavenumbers larger than, say, $k_0\sim 1/0.1R$. 
This means that the 'Kolmogorov-like' behavior in the autocorrelation function is typically at the part corresponding to shifts $\ell\lesssim 1/k_0$. 
The autocorrelation at larger $\ell$ and the spectra at smaller $k$ could contain information about the injection of the turbulent energy and the supernova explosion. 
This is in agreement with the Kolmogorov theory. He aimed at description of the \textit{local} structure of turbulence, i.e., on the length scales much smaller than the largest shift. Therefore, it is natural to expect deviation from the power law on scales comparable to the size of the object because they are not 'local enough' for the domain limited by the size of an object.  

Note that the continuous Fourier transform (CFT) should be used to determine length-scales of the fluctuations (related either to their periodicity or to the sizes). This is especially important for the small $k$: there are modes in the spectrum located at $k$ that are related to the non-integer fractions of the SNR radius, while FFT returns spectra for the integer wavenumbers only. 

There are a few ways to calculate the average within a sector. In Fig.~\ref{turb:fig-effect-wiener}, we compare the average autocorrelation function $\overline{C_{\delta {\cal I}}}(\ell)$ (solid line) and the autocorrelation calculated from the average profile of brightness $C_{\overline{\delta {\cal I}}}(\ell)$ (black dashed line) where the overline marks the average over the 45 individual functions or brightness profiles within the sector. There are also power spectra calculated similarly: the average $\overline{{\cal F}\left\{{\cal I}\right\}^2}(k)$ and from the average profile ${\cal F}\left\{\overline{{\cal I}}\right\}^2(k)$. There is an interesting feature, besides the overall similarity and local differences between the lines. Namely, the shape and location of the first few peaks in the spectra are the same for solid and dashed lines. They should correspond to some features in the brightness profiles and be reflected in the autocorrelation function. Note that they may be related either to an HD property of SNR or to the disturbances. 
In particular, the second and third modes in the X-ray spectrum are at wavenumbers corresponding to the length scales $0.39R$ and $0.27R$. There is a tiny local maximum at $\ell=0.27R$ in the X-ray autocorrelation functions shown by the solid line in Fig.~\ref{turb:fig-effect-wiener}. The size of a corresponding feature in the brightness profile should be about $1'$ in the radial direction. It is likely related to the thickness of the maximum in ${\cal I}\rs{x}$ in between $(0.55\div 0.8)R$. It is due to a projection effect of a shell-like object onto the plane of the sky and, therefore, is not related to the turbulence. The power spectrum mode related to $0.37R$ could be due to a periodicity in the brightness profile, e.g., between the two peaks around $r\approx 0.05R$ and $0.4R$. Note that there is a decrease in error bars at $\ell=0.39R$ in the autocorrelation function, which could be a sign that this feature is related to a well-defined periodicity in the surface brightness profile.

\section{Conclusions}

We performed autocorrelation and power spectrum analyses of the X-ray and radio images of the Tycho SNR. We determined the Kolmogorov-like turbulence in this SNR, with the power spectrum index $8/3$ for 2D and $5/3$ for 1D analysis. This is in agreement with the findings by \citet{2018MNRAS.480.2200S} obtained from 1D azimuthal variations of the radio brightness in the Tycho SNR. 
Interestingly, one of the earliest observational evidence about Kolmogorov spectrum in astronomical structures was found by \citet{1955AZh....32..255K}, while he was working in our city Lviv \citep[see Sect.~5 in][]{2022EPJH...47...12N}.

The 2D X-ray and radio autocorrelation functions reveal anisotropies aligned with the gradients in density and magnetic field in the ISM surrounding the SNR. 

The slope of the power spectra appears to show only weak sensitivity to the line-of-sight integration. 
Generally, the observed power spectrum from a 2D image of a 3D object represents the sum of local power spectra from a sequence of 2D cross-sections perpendicular to the line of sight. If each of these local spectra has a different shape than the observed spectrum, it would have a certain slope (e.g., Kolmogorov’s 8/3) rather by chance. That is unlikely, because we derived the Kolmogorov spectra for different X-ray photon energy ranges, for X-rays and radio bands, for different contours over the SNR, and for 2D and 1D brightness distributions. All of these together point to the idea that fully developed turbulence is present everywhere throughout the Tycho remnant as a 3D structure. If so, then the local spectra at each cross-section along LoS are also of the Kolmogorov type. A superposition of spectra sharing the same slope, regardless of their normalization, also exhibits this slope value.

We analyzed fluctuations in surface brightness along 1D azimuthal contours in the X-ray and radio images. Our results confirm the findings of \citet{2018MNRAS.480.2200S}, showing that the structure function closely follows a power-law dependence $\ell^{2/3}$, even though our contour differs from the one used in the refereed paper. 
We note that the original Kolmogorov theory was developed for radial variations of a parameter. In principle, it should be demonstrated that the $2/3$ slope also applies to azimuthal variations. However, since Kolmogorov's theory assumes local randomness of a physical parameter, we hypothesize that the turbulence spectrum may still follow a Kolmogorov-like form (within some broad range of wavenumbers) along any sufficiently smooth contour or curve on the SNR projection.

There are two distinct types of 1D autocorrelation functions and power spectra that can be obtained from an SNR image. 
The first may be derived from the 2D autocorrelation function shown in Fig.~\ref{turb:fig-2Dautocor}, which is calculated for the shifts of the whole original 2D SNR image in different directions, by averaging the radial distributions from that figure over the azimuth. The power spectrum relevant for such an average autocorrelation has a slope close to $8/3$, which corresponds to the 2D Kolmogorov turbulence. The second type of 1D analysis is relevant to the procedure considered by \citet{1941DoSSR..30..301K}, where the properties of fluctuations were considered along radial directions of a higher-dimensional object.
We divided an image of Tycho SNR into a number of sectors. Our analysis 
shows that the power spectra of fluctuations follow a Kolmogorov-like $k^{-5/3}$ law across all directions and a broad range of $k$. These results confirm the presence of fully developed turbulence and reveal some sector-dependent features, likely linked to periodicity in structures or their sizes, which could not be related to the turbulence but emerge due to a projection effect. 

The slope of the power spectrum, such as the Kolmogorov $5/3$ slope, typically emerges for eddies in a plasma under steady-state conditions, without significant energy losses or inflow. However, SNR is not a steady-state system. The fact that fully developed turbulence is observed in Tycho’s SNR implies that the timescale to establish a stationary cascade within the inertial range is shorter than the age of the remnant. 
It would be valuable to study the \textit{temporal} evolution of turbulence from the moment of the supernova explosion to the point when Kolmogorov-type behavior emerges. Investigating the \textit{spatial} development of turbulence could clarify the dominant physical drivers and characteristic scales involved, in other words, whether turbulence in SNRs is primarily driven by the shock or by structures injected during the explosion itself.

In this context, we emphasize the importance of high-resolution imaging of SNRs for studying the turbulence. Instruments such as Chandra and the VLA enable the analysis of fluctuations across a broad range of spatial scales, from the SNR radius $R$ down to $\sim 0.01R$. However, these observations do not provide information about turbulence on scales much smaller than $0.01R$, which are relevant for processes of particle acceleration at shocks. This limitation arises due to change of the power spectrum at wavenumbers corresponding to $\sim 1\%$ of the SNR radius, a feature that we believe is of an instrumental origin.

The CR-driven instabilities may modify the spectra of turbulence at smaller scales. Indeed, the length-scale which corresponds to the fastest-growth mode of the \citet{2004MNRAS.353..550B} instability is 
$ \ell\rs{Bell}=2\pi/k\rs{max} \sim 0.1\,V_{3500}^{-3}n_{0.1}^{-1}B_{3}E_{100}\,\un{pc}$ 
where the parameters are in units relevant to Tycho SNR: the shock velocity in 3500 km/s, the interstellar number density in $0.1 \un{cm^{-3}}$, magnetic field strength in $3\un{\mu G}$, and the CR maximum energy in 100 TeV. This $\ell\rs{Bell}$ is a fraction $\sim 0.03$ of the Tycho SNR radius, which is comparable to the limiting length-scale $3''$ ($\approx 0.01R$), to which we can study the power spectrum in the Chandra images of Tycho SNR (Appendix~\ref{turb-app-fon}):  spectrum of fluctuations in the observed image is contaminated by the instrumental and background noise from wavenumbers corresponding to such scales. 

The instruments which provide a reliable power spectrum in the SNR image on scales around and below $0.01R$ could advance our knowledge about the spectra of CR-driven instabilities, resonant or non-resonant \citep{2024ApJ...967...71Z}, or about effects of CR transport through MHD turbulence \citep{2020ApJ...894...63X,2022ApJ...927...94X}.

\begin{acknowledgments}
We thank the anonymous referee for a constructive report which helped us to improve the manuscript.
This project has received funding through the MSCA4Ukraine project, which is funded by the European Union. Views and opinions expressed are however those of the author(s) only and do not necessarily reflect those of the European Union. Neither the European Union nor the MSCA4Ukraine Consortium as a whole nor any individual member institutions of the MSCA4Ukraine Consortium can be held responsible for them.
We also acknowledge the support from INAF 2023 RS4 Theory grant and INAF 2023 RS4 Mini grant. 
We thank the Armed Forces of Ukraine for providing security to perform this work. 
This paper employs a list of Chandra datasets, obtained by the Chandra X-ray Observatory, contained in~\dataset[DOI: 10.25574/cdc.259]{https://doi.org/10.25574/cdc.259}. This work made use of the HPC cluster at DESY, Zeuthen, Germany.
\end{acknowledgments}

\appendix

\section{Cancellation of the background noise}
\label{turb-app-fon}

\begin{figure*}
  \centering 
  \includegraphics[width=0.6\textwidth]{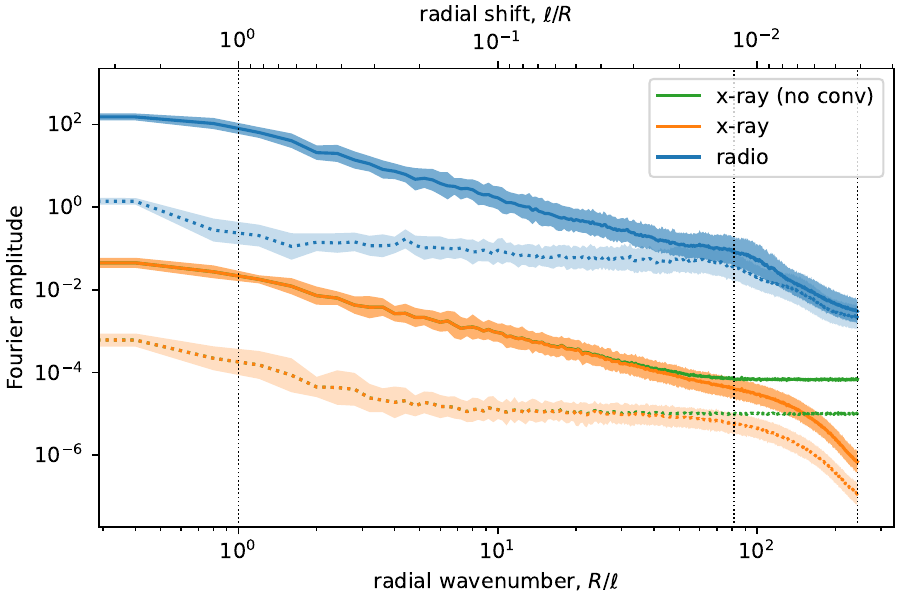}
  \caption{Azimuthal average of the radial component of Fourier amplitudes from 2D Fourier transform of the X-ray image of Tycho SNR (orange and green lines) and its radio image (blue lines). Solid lines represent the FFT of the image, while dashed lines represent the FFT of the background noise. The vertical lines correspond to the average radius $R=4'$ (left), to the wavenumber $k=1/3''$ (middle), and the Nyquist wavenumber (right). Green lines correspond to the X-ray data before, and the orange lines after the Gaussian smoothing with a 1-pixel kernel was applied to the X-ray image. The green lines coincide with the orange lines at wavenumbers $R/\ell\lesssim 30$.
   Shaded area shows $1\sigma$ errors for the log Gaussian; the errors are statistical, due to averaging between the radial profiles in 2D Fourier space at azimuths from 0 to $2\pi$.%
  }
  \label{turb:fig-radial-fft}
\end{figure*}

The common procedure for eliminating background noise consists of a few steps. We perform a 2D FFT of the observed 2D image and of a nearby region outside the SNR, which has the same size as the observed image. Then the noise is removed by adding it in anti-phase to the Fourier transform of the SNR image in the Fourier domain. An inverse Fourier transform then gives the cleaned-up image of the SNR. The Fourier spectra averaged over the azimuth are shown in Fig.~\ref{turb:fig-radial-fft}. One can see that the noise in the X-ray image is much lower than the signal at all length scales. 

\begin{figure*}
  \centering 
  \includegraphics[width=0.85\textwidth]{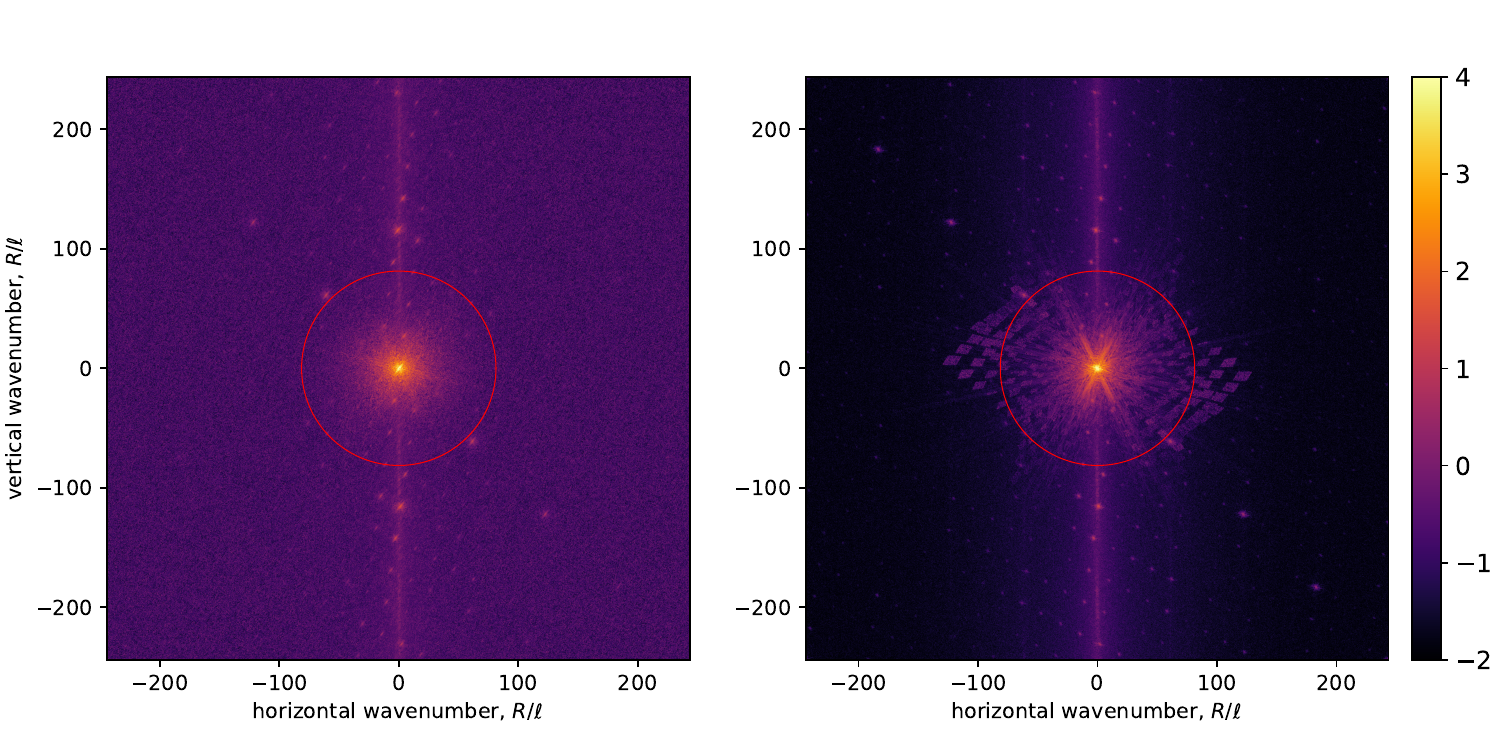}
  \caption{2D discrete Fourier transform of the autocorrelation function from the image before the background noise cancellation,  
   for the X-ray image of Tycho SNR (left column) and its radio image (right column). The color scale corresponds to the decimal logarithm of the Fourier amplitudes. The red circle marks the wavenumber $1/3''$. The same distributions after the background noise cancellation are shown in Fig.~\ref{turb:fig-2Dfourier}.%
  }
  \label{turb:ap:fig-2d-autocorr-fft}
\end{figure*}

There is a feature in the X-ray Fourier spectrum: it behaves like white noise above $R/\ell > 80$, which corresponds to wavenumbers $k>1/3''$. 
Such a feature seems to be an instrumental effect, probably due to the scattering of X-ray photons on length scales of a few pixels in the detector. It could arise from a combination of dithering during the observation, random patterns in the PSF off-axis morphology, and the shape of the PSF. In fact, about 40\% of the on-axis photons are spread out of a pixel, while 95\% of them fall within a 2-pixel radius\footnote{\url{https://cxc.cfa.harvard.edu/ciao/PSFs/psf_central.html}}. To reduce this effect, we applied Gaussian smoothing to the initial X-ray image with a 1-pixel kernel that suppresses this feature (cf. the green and orange lines on Fig.~\ref{turb:fig-radial-fft}). Then we apply the noise-canceling procedure described above to the filtered X-ray data. The resulting azimuth-averaged spectrum is shown in Fig.~\ref{turb:fig-radial-fft}. 
Fig.~\ref{turb:ap:fig-2d-autocorr-fft} shows the 2D power spectra for the original images with background noise. 
The power spectrum of the images after filtering off the instrumental effect and the background noise cancellation is shown in Fig.~\ref{turb:fig-2Dfourier}.


\begin{table}
\centering
\caption{Parameters calculated from the images of Tycho SNR after the noise cancellation. The numbers in the row marked 'total' correspond to the whole SNR image.}
\begin{tabular}{ccccccc}
\hline\hline
$r/{\cal R}$ &$\overline{\cal I}$ &$\sigma_{\cal I}$ &$\overline{a}$    &$\sigma_{a}$ &$c_1$\\
\hline\hline
\multicolumn{6}{c}{X-day data}\\
\hline
0.6	&1.97e-07	&8.91e-08	&4.34e-04	&9.70e-05	&0.667\\
0.7	&2.79e-07	&1.41e-07	&5.11e-04	&1.33e-04	&0.737\\
0.8	&3.51e-07	&2.22e-07	&5.67e-04	&1.73e-04	&0.666\\
0.9	&2.48e-07	&1.31e-07	&4.81e-04	&1.31e-04	&0.756\\
0.95	&1.60e-07	&1.30e-07	&3.76e-04	&1.36e-04	&0.582\\
total &2.13e-07 &1.67e-07 &4.25e-04 &1.81e-04 &0.863\\
\hline\hline
\multicolumn{6}{c}{radio data}\\
\hline
0.6	&7.70e-04	&1.51e-04	&2.76e-02	&2.73e-03	&0.602\\
0.7	&9.61e-04	&2.11e-04	&3.08e-02	&3.41e-03	&0.613\\
0.8	&1.19e-03	&3.04e-04	&3.41e-02	&4.56e-03	&0.672\\
0.9	&1.13e-03	&3.60e-04	&3.33e-02	&5.30e-03	&0.645\\
0.95	&9.73e-04	&3.30e-04	&3.07e-02	&5.39e-03	&0.697\\
total & 8.30e-04 & 3.73e-04 & 2.80e-02 & 6.66e-03 & 0.764 \\
\hline
\hline
\end{tabular}
\label{turb-app-tabl1}
\end{table}

\begin{figure*}
  \centering 
  \includegraphics[width=0.94\textwidth]{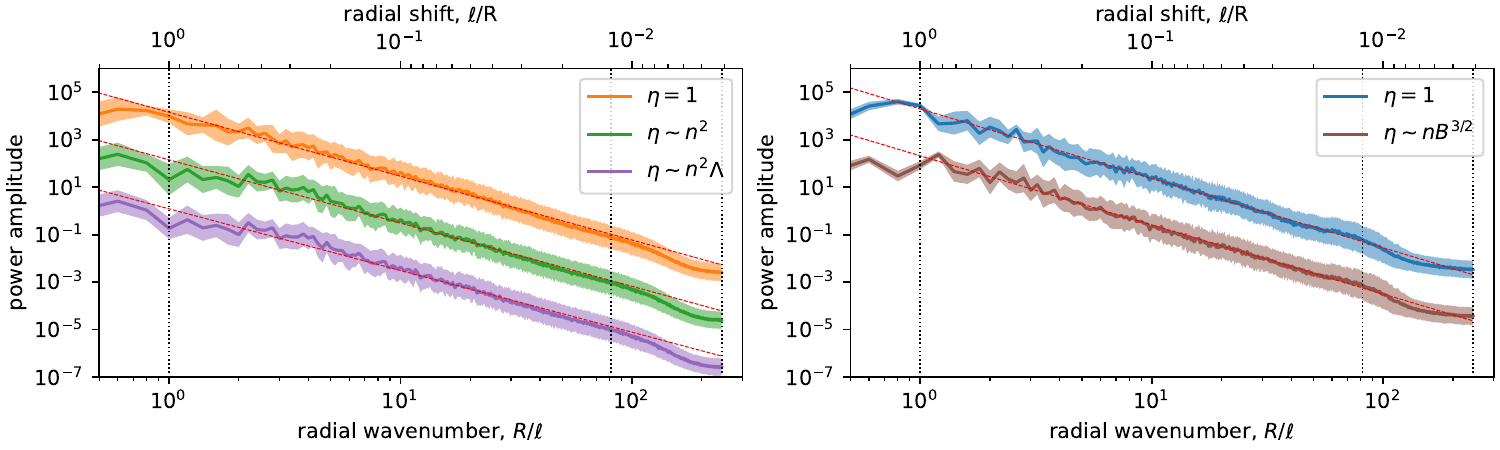}
  \caption{%
    The power spectra of fluctuations in the X-ray (left) and radio (right) images of the Tycho SNR calculated as FFT amplitudes of the autocorrelation functions and averaged over the azimuthal angle. The spectra are obtained after applying a correction for a geometrical factor for each pixel in the image. Different assumptions about the geometrical factor $\eta$ are considered. The shaded region displays $1\sigma$ statistical uncertainty. To prevent overlapping, consecutive spectra were shifted by a factor of $1/100$. The red dashed line corresponds to the best-fit power laws; see the text for their spectral indices.%
  }
  \label{turb:ap:fig-power-spectrum-with-eta}
\end{figure*}
\begin{figure*}
  \centering 
  \includegraphics[width=0.94\textwidth]{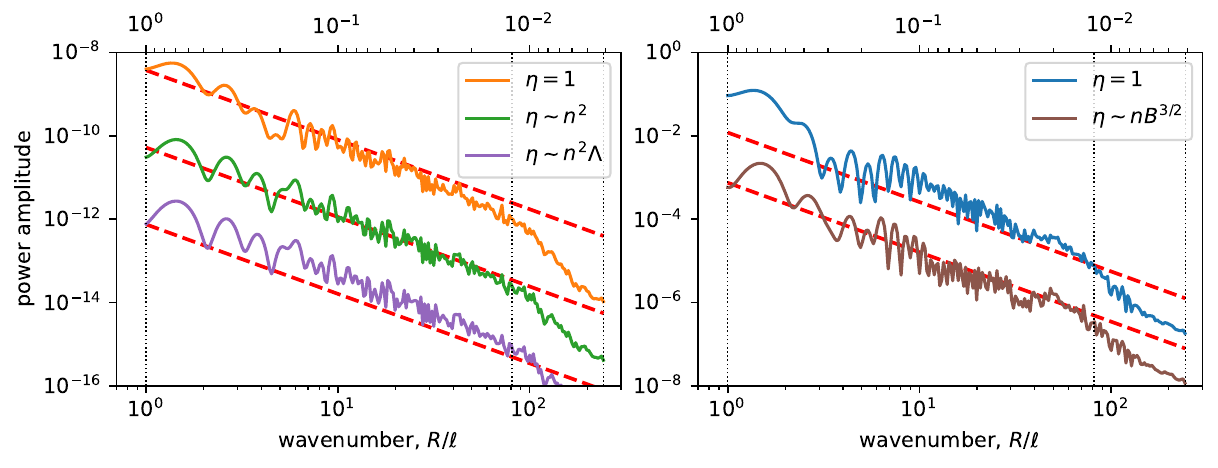}
  \caption{%
    The same power spectra as shown by the solid lines in Fig.~\ref{turb:fig-effect-wiener}, for X-rays (left) and radio (right). Different assumptions about the geometrical factor $\eta$ are considered. Red dashed lines correspond to $k^{-5/3}$.%
  }
  \label{turb:ap:fig-sectors-xray-radio-fft-eta}
\end{figure*}

\section{Uncertainties}
\label{turb-app-errors}

\subsection{Geometrical factor}
\label{turb-app-geom-factor}

The geometrical factors account for the variation of the parameters along the line of sight in the SNR interior. They are calculated for the X-ray and radio images by taking into account the 3D spatial distributions of MHD parameters in a remnant of SN type Ia at the age of 450 years obtained in numerical simulations \citep[for more details see][]{2024ApJ...972...63P}. 

Fig.~\ref{turb:ap:fig-power-spectrum-with-eta} shows the same power spectra as in the top plot in Fig.~\ref{turb:fig-1Dfourier}, calculated with different geometrical factors. In particular, the orange and blue lines on Fig.~\ref{turb:ap:fig-power-spectrum-with-eta} are the same as in the top plot in Fig.~\ref{turb:fig-1Dfourier}. Dashed black line on Fig.~\ref{turb:fig-1Dfourier} corresponds to the green line in Fig.~\ref{turb:ap:fig-power-spectrum-with-eta}. It is calculated from the X-ray image corrected for the geometrical factor, which accounts for the spatial distribution of $n^2$ in the SNR interior along the line of sight. The violet line in Fig.~\ref{turb:ap:fig-power-spectrum-with-eta} is calculated with the geometrical factor that includes the spatial variations of $n^2\Lambda(T,\tau)$ where the cooling function $\Lambda(T,\tau)$ depends on the internal structure of the temperature $T$ and the ionization parameter $\tau=n\rs{e}t$. The brown line considers the distribution of $nB^{3/2}$. 

The red dashed lines in Fig.~\ref{turb:ap:fig-power-spectrum-with-eta} show the best-fits by a linear function of the spectra in log-log space. The fits were performed for the wavenumbers $1\div 80$, i.e., within the vertical lines. 
The values of the spectral indices for these best-fit lines are: 
$-2.69\pm0.07$ (orange), $-2.67\pm0.08$ (green), $-2.60\pm0.08$ (violet), $-2.93\pm 0.06$ (blue), $-2.91\pm0.06$ (brown). By comparing to the spectral index for the Kolmogorov turbulence $8/3=2.667$, we may say that  
i) inclusion of the geometrical factor into calculations of the power spectrum does not change essentially the slope of the spectrum; ii) the spectral index of the turbulence in the X-ray image is well comparable, within errors, with $8/3$ value even for $\eta\rs{x}=1$; iii) the spectral index for the radio image is slightly larger than $8/3$ but may be closer to this value in a narrower range of $k$; iv) consideration of the ionization parameter and temperature does not make essential improvement in the value of the spectral index of turbulence in the X-ray image in energy $1.2-4.0\un{keV}$. The reason for the last point is that, initially, we have chosen the X-ray photon energy range 1.2-4.0 keV in which the cooling function is almost constant for various temperatures and ionization states of plasma \citep[see Fig.5 in][]{2024ApJ...972...63P}. 

The above statements are related to the azimuthal average of the 2D power spectrum. What could be the effect of the geometrical factor on the spectra from 1D profiles of brightness, where the procedure of averaging does not smooth out detailed features? Fig.~\ref{turb:ap:fig-sectors-xray-radio-fft-eta} demonstrates such spectra. The orange and blue lines there correspond to the same lines in Fig.~\ref{turb:fig-effect-wiener}. Other lines include different approaches to the geometric factor. As in the previous case, the inclusion of $\eta$ does not alter the conclusion about the type of turbulence.

\subsection{Other sources of uncertainties}

Besides the lack of knowledge about the structure of the SNR along the line of sight (LoS), errors in our analysis could (i) arise from background noise, (ii) be of instrumental origin, or (iii) result from averaging the data. The observational errors are determined by the first two factors.

In order to understand the role of the background emission in our analysis, we compare the power spectra from the source and the background region of a similar size and perform the procedure of cancellation of the noise which is not related to the source, as described in Appendix~\ref{turb-app-fon}. It appears that the Fourier spectrum of the X-ray background is orders of magnitude below the spectrum from the source, over all $k$. It is also negligible for the radio data, at $k\lesssim 1/0.02R$ (Fig.~\ref{turb:fig-radial-fft}). Nevertheless, we subtract the background in the Fourier space for all $k$ both in the X-ray and in the radio data. 

The instrumental errors in X-rays are visible from comparison of the green and orange lines in Fig.~\ref{turb:fig-radial-fft}. In order to get rid of the instrumental effect, we apply a convolution to the images with a small-kernel Gaussian. Such a procedure resembles a high-frequency filter. It lowers the instrumental 'noise' which dominates the spectra at high $k\gtrsim 1/0.02R$. Note that destruction of the power spectrum of fluctuations in the radio image by to background emission is prominent at the same $k\gtrsim 1/0.02R$. Therefore, we limit our analysis of the power spectra to wavenumbers smaller than $k\sim 1/0.02R$. The power spectra extracted from observations are highly uncertain at larger $k$.

Typical statistical errors appearing in the course of azimuthal averaging of the power spectra are shown, e.g., in Figs~\ref{turb:fig-radial-fft} and \ref{turb:ap:fig-power-spectrum-with-eta}. At these plots, they are $1\sigma$ intervals for a log Gaussian and demonstrate the stability of determination of the power-law index for the power spectra. Azimuthal averaging lowers fluctuations between different wavenumbers in the power spectra that are present in the spectra for individual azimuthal directions (compare, for example, Figs~\ref{turb:ap:fig-power-spectrum-with-eta} and \ref{turb:ap:fig-sectors-xray-radio-fft-eta}). In order to lower the random fluctuations in the form of the white noise over the power spectra, in some cases we applied the Wiener filter \citep[][sect.~9.2.4]{1990ph...book.....L}. Such a procedure preserves the main dependence while making the shape of the spectrum more clearly visible.

\section{Derivation of equations for the density fluctuations}
\label{turb-app-deltan-derivation}

We follow here the approach by \citet{2018MNRAS.480.2200S,2022FrASS...9.2467S} to derive equations (\ref{turb:eq-structfunc}) and (\ref{turbeq:eq-corr-n}).

Let us denote $a={\cal I}^{1/2}$, $b={\cal I}'^{1/2}$ and use the equality
$2a^2b^2=a^4+b^4-(a+b)^2(a-b)^2$. Then 
\begin{equation}
 C_{\cal I}(\ell)\equiv
 \langle {\cal I}{\cal I}'\rangle=\frac{1}{2}\langle {\cal I}{\cal I}\rangle+\frac{1}{2}\langle {\cal I}'{\cal I}'\rangle-
 \frac{1}{2}\left\langle \left(a+b\right)^2\left(a-b\right)^2\right\rangle.
 \label{turb-ap1eq1}
\end{equation}
We denote the first two terms on the right side by $C_1\equiv C_{\cal I}(\ell\rs{min})$. 
By introducing $a=\bar a+\delta a$ and $b=\bar b+\delta b$ and noting that $\bar a=\bar b$, 
we re-write (\ref{turb-ap1eq1}) as 
\begin{equation}
 C_{\cal I}(\ell)=C_1-
 2{\bar a}^2\left\langle \left(1+\frac{\delta a}{2\bar a}+\frac{\delta b}{2\bar b}\right)^2\left(\delta a-\delta b\right)^2\right\rangle.
\end{equation}
The two ratios can be estimated as $\delta a/\bar a\simeq \sigma_a/\bar a$ and $\delta b/\bar b\simeq \sigma_b/\bar b$ with $\sigma_a=\sigma_b$. Then
\begin{equation}
 C_{\cal I}(\ell)=C_1-
 c_1\left\langle \left(\delta a-\delta b\right)^2\right\rangle_a,
 \qquad
 c_1=2{\bar a}^2\left(1+{\sigma_a}/{\bar a}\right)^2\cdot \sigma_a^2/\sigma_{\cal I}^2.
 \label{turn-app-eq3}
\end{equation}
where $c_1$ is a dimensionless constant which may be calculated from the observed images of the Tycho SNR (Table~\ref{turb-app-tabl1}). 
Note that the operator $\langle\rangle_a\propto \sigma_a^{-2}$ while $\langle\rangle\propto \sigma_{\cal I}^{-2}$. At this point, we may use the approximate relations $a\propto n$ and $b\propto n'$ to write 
\begin{equation}
 \left\langle \left(\delta a-\delta b\right)^2\right\rangle_a\approx\left\langle \left(\delta n-\delta n'\right)^2\right\rangle_n
 \equiv \frac{1}{\sigma_{n}^2 N}{\sum\limits_{(i,j)}^{N}\left(\delta n-\delta n'\right)^2}.
 \label{turn-app-eq5}
\end{equation}
This transforms the equation (\ref{turn-app-eq3}) into (\ref{turb:eq-structfunc}).

We may further transform (\ref{turn-app-eq3}) by writing 
$\langle (\delta a-\delta b)^2\rangle_a=
\langle \delta a^2+\delta b^2-2\delta a\delta b\rangle_a\approx
C_2-2\langle\delta n\cdot\delta n'\rangle_n$
where $C_2$ is defined as $C_2= \langle \delta a \cdot\delta a\rangle_a+\langle \delta b\cdot \delta b\rangle_b$ and it should be $C_2=2$ because the autocorrelation in $C_2$ involves deviations from the average. Then we have an expression for the autocorrelation of the plasma density fluctuations:
\begin{equation}
 C_{\cal I}(\ell)\approx C_1
 +2c_1\big(\langle\delta n\cdot\delta n'\rangle_n-1\big).
 \label{turn-app-eq4}
\end{equation}
At this point, it is straightforward to see that
\begin{equation}
\left\langle\delta n\cdot\delta n'\right\rangle_n=
1-\frac{1}{2}\left\langle\left(\delta n-\delta n'\right)^2\right\rangle_n.
 \label{turb-app-eq6}
\end{equation}
The Fourier transform of $\left\langle\delta n\cdot\delta n'\right\rangle_n$ yields the power spectrum. This represents the relationship between the structure function and the power spectrum.

\section{Radial autocorrelation functions and spectra for Tycho SNR}
\label{turb-app-1D-radial-autocorr}

\begin{figure*}
  \centering 
  \includegraphics[width=0.94\textwidth]{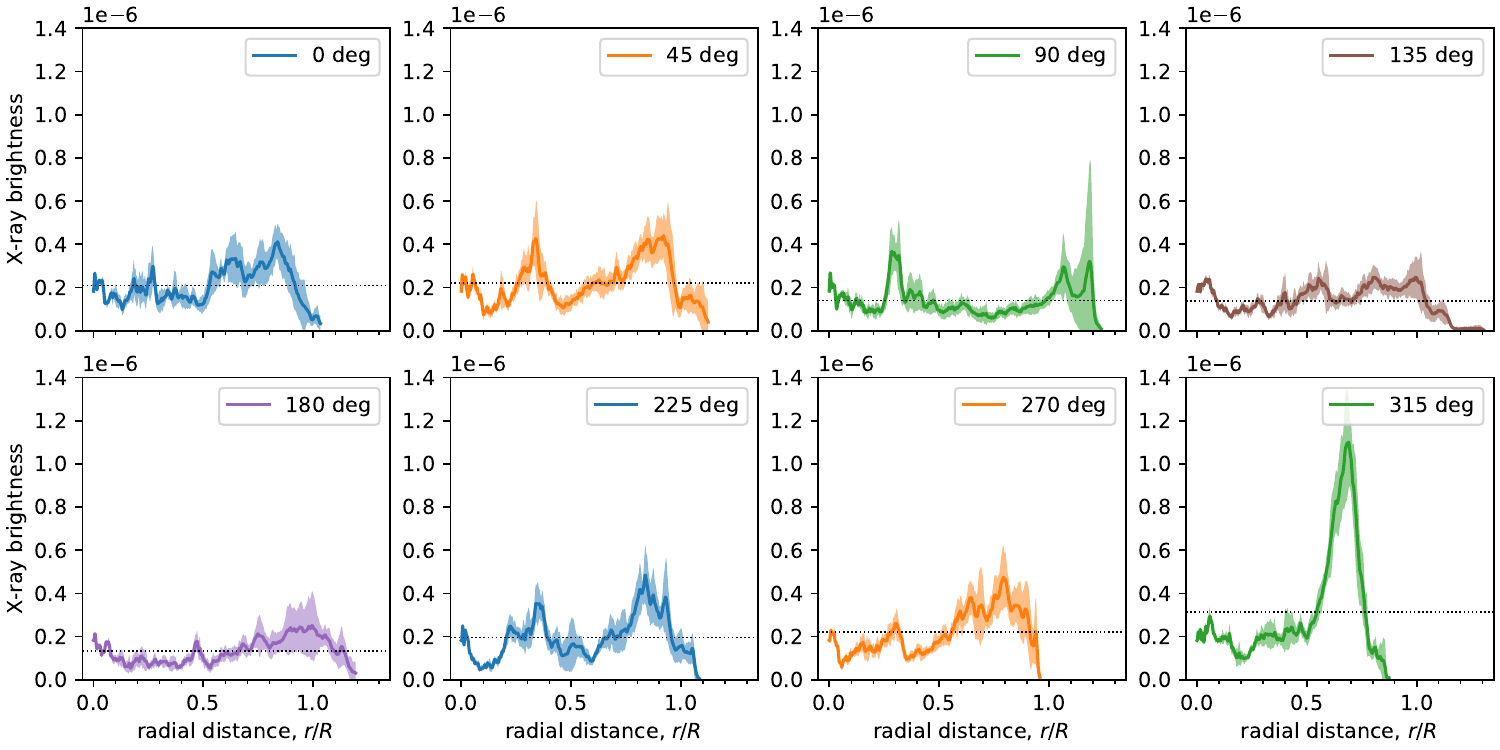}
  \caption{%
    Radial profiles extracted from the X-ray image of Tycho SNR. Profiles are calculated as an average within $15^\circ$ sectors, which begin at the angle written in the legend of each plot. Dotted horizontal lines mark the profile's average value. The errors shown are at the $1\sigma$ level.%
  }
  \label{turb:ap:fig-sectors-profiles-xray}
\end{figure*}

\begin{figure*}
  \centering 
  \includegraphics[width=0.94\textwidth]{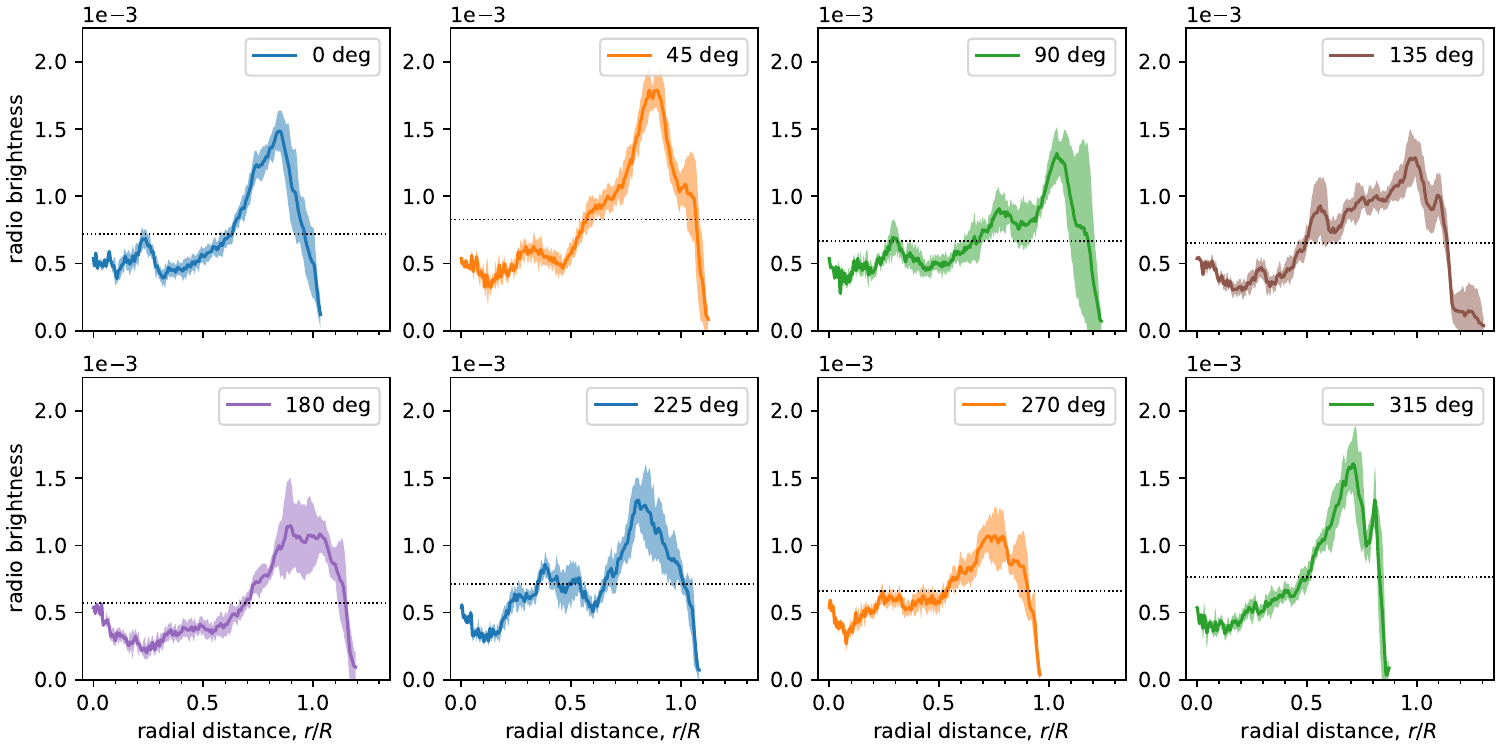}
  \caption{%
  Same as in Fig.~\ref{turb:ap:fig-sectors-profiles-xray} for the radio image of Tycho SNR.%
  }
  \label{turb:ap:fig-sectors-profiles-radio}
\end{figure*}

To visualize the differences in properties of disturbances in the X-ray and radio images of Tycho SNR, we present a set of relevant plots here. The images (Fig.~\ref{turb:fig-images}) are divided into $15^{\circ}$ sectors, with a center at the explosion point. Within each sector, we took $45$ radial profiles of surface brightness ${\cal I}$ and calculated their averages. The average profiles with $1\sigma$ errors for a number of directions are shown in Fig.~\ref{turb:ap:fig-sectors-profiles-xray} for X-ray and Fig.~\ref{turb:ap:fig-sectors-profiles-radio} for the radio band. 

\begin{figure*}
  \centering 
  \includegraphics[width=0.94\textwidth]{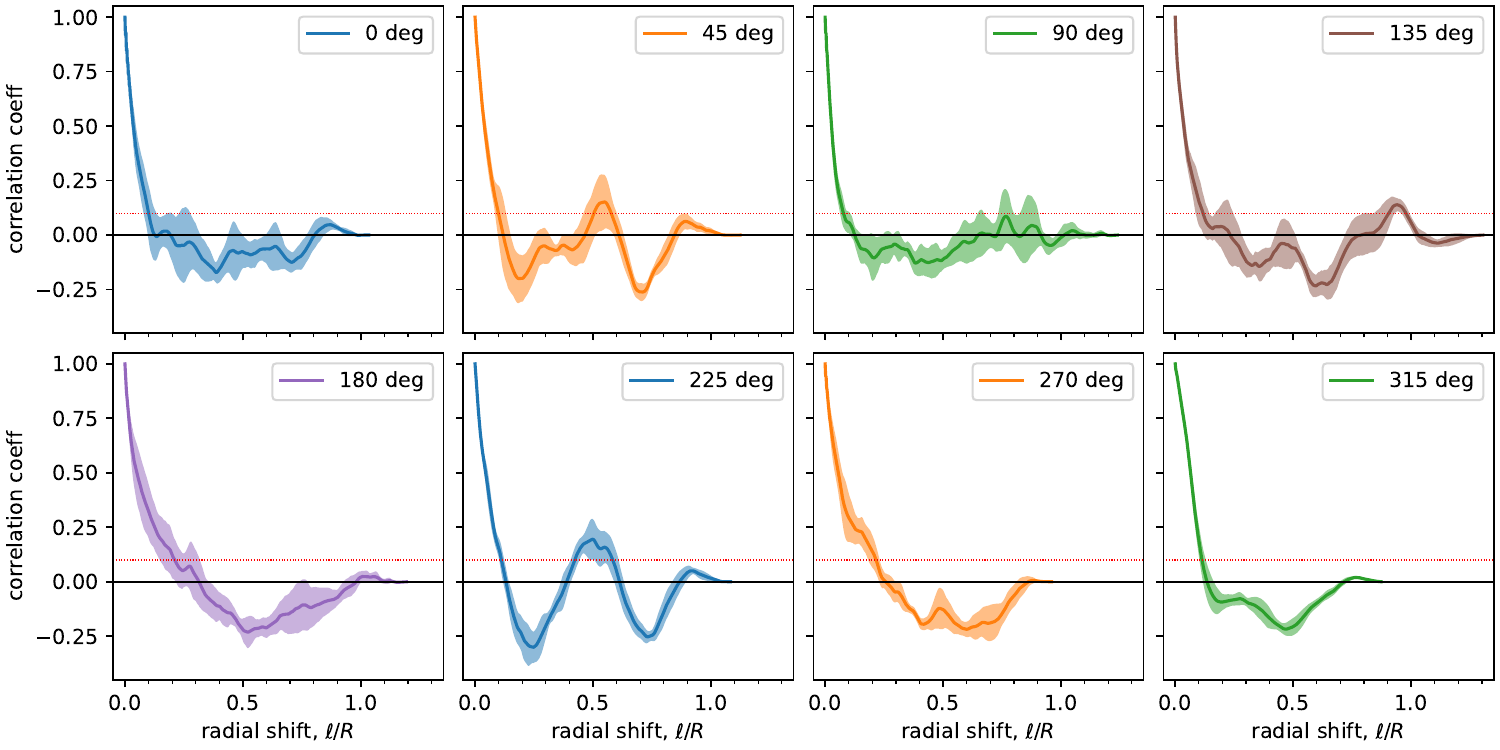}
  \caption{%
    Radial autocorrelation functions from the radial profiles extracted from the X-ray image of Tycho SNR. They are calculated for each of the individual profiles within a $15^\circ$ sector and then averaged across the sector. The numbers in the legend correspond to the angle where a given sector begins. 
    The errors shown are at the $1\sigma$ level. The red dashed line is at the correlation of 0.1.%
  }
  \label{turb:ap:fig-sectors-autocorr-xray}
\end{figure*}

\begin{figure*}
  \centering 
  \includegraphics[width=0.94\textwidth]{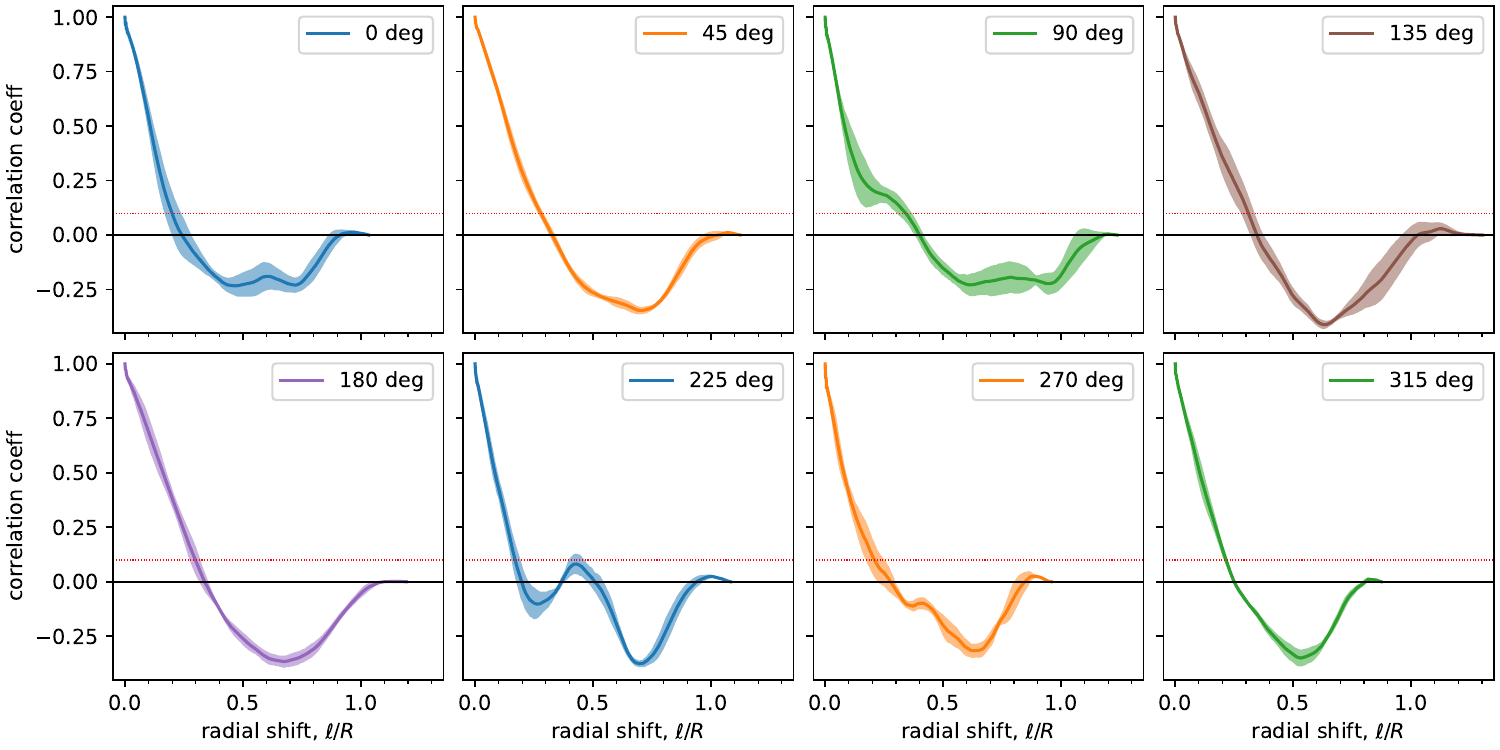}
  \caption{%
  Same as in Fig.~\ref{turb:ap:fig-sectors-autocorr-xray} but for the radio image.%
  }
  \label{turb:ap:fig-sectors-autocorr-radio}
\end{figure*}

In a similar fashion, we calculated the autocorrelation functions $C_{\cal \delta I}(\ell)$ for each individual radial brightness profile within a sector and then the average between them. Each of the 45 profiles was shifted in its own radial direction. The average autocorrelation functions for a number of sectors are shown in Fig.~\ref{turb:ap:fig-sectors-autocorr-xray} and Fig.~\ref{turb:ap:fig-sectors-autocorr-radio} for X-ray and radio images, respectively.  

\begin{figure*}
  \centering 
  \includegraphics[width=0.94\textwidth]{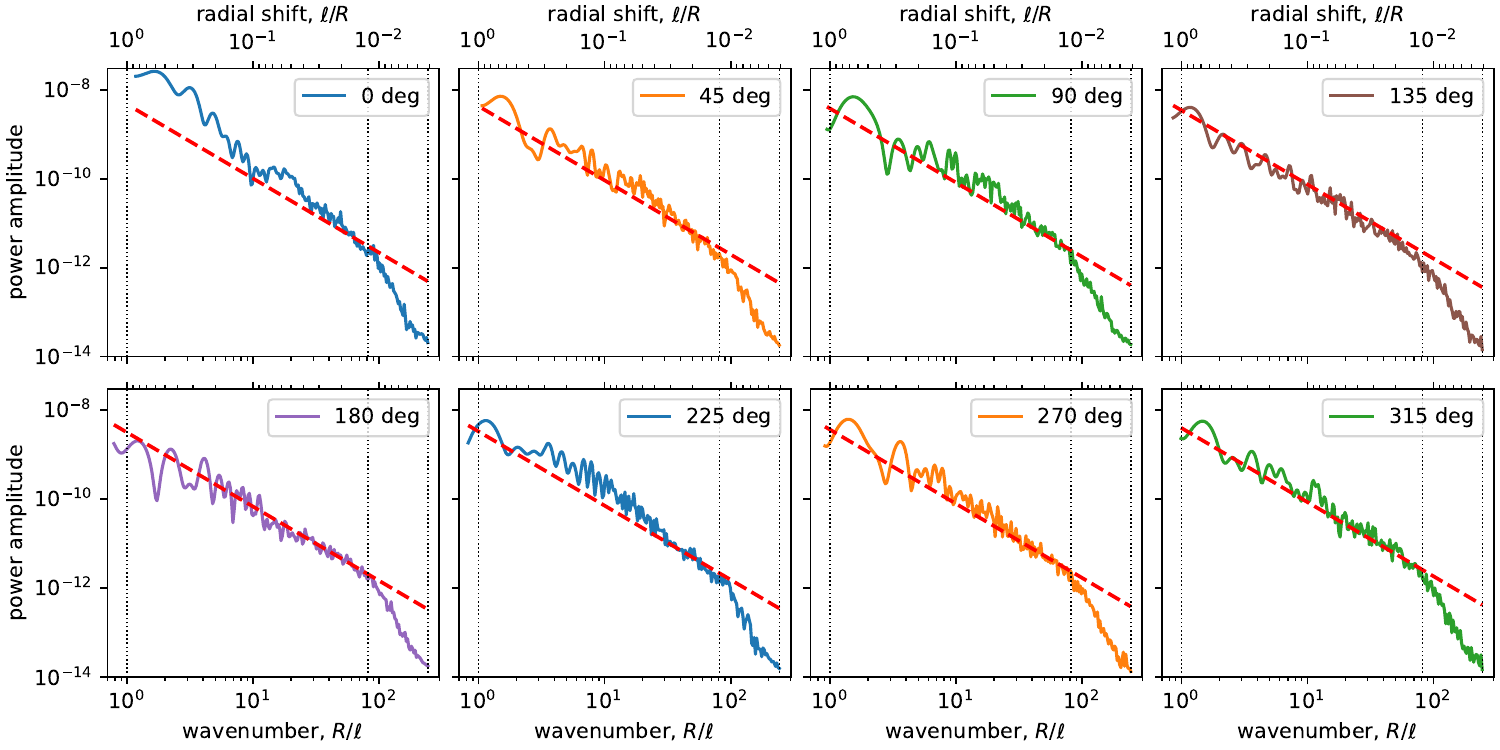}
  \caption{%
    The average power spectra for a number of sectors calculated from continuous Fourier transforms of the individual radial profiles of X-ray brightness within the sector. The red dashed lines indicate the Kolmogorov-type $5/3$ power law, the amplitude is the same on all plots. 
    Spatial scales are normalized to the average radius $R = 4'$.
    Vertical lines mark the wavenumbers $1/R$, $1/3''$, and the Nyquist one. Spectra start at different $k$ because the radius of each sector is different.%
  }
  \label{turb:ap:fig-sectors-xray-fft}
\end{figure*}

\begin{figure*}
  \centering 
  \includegraphics[width=0.94\textwidth]{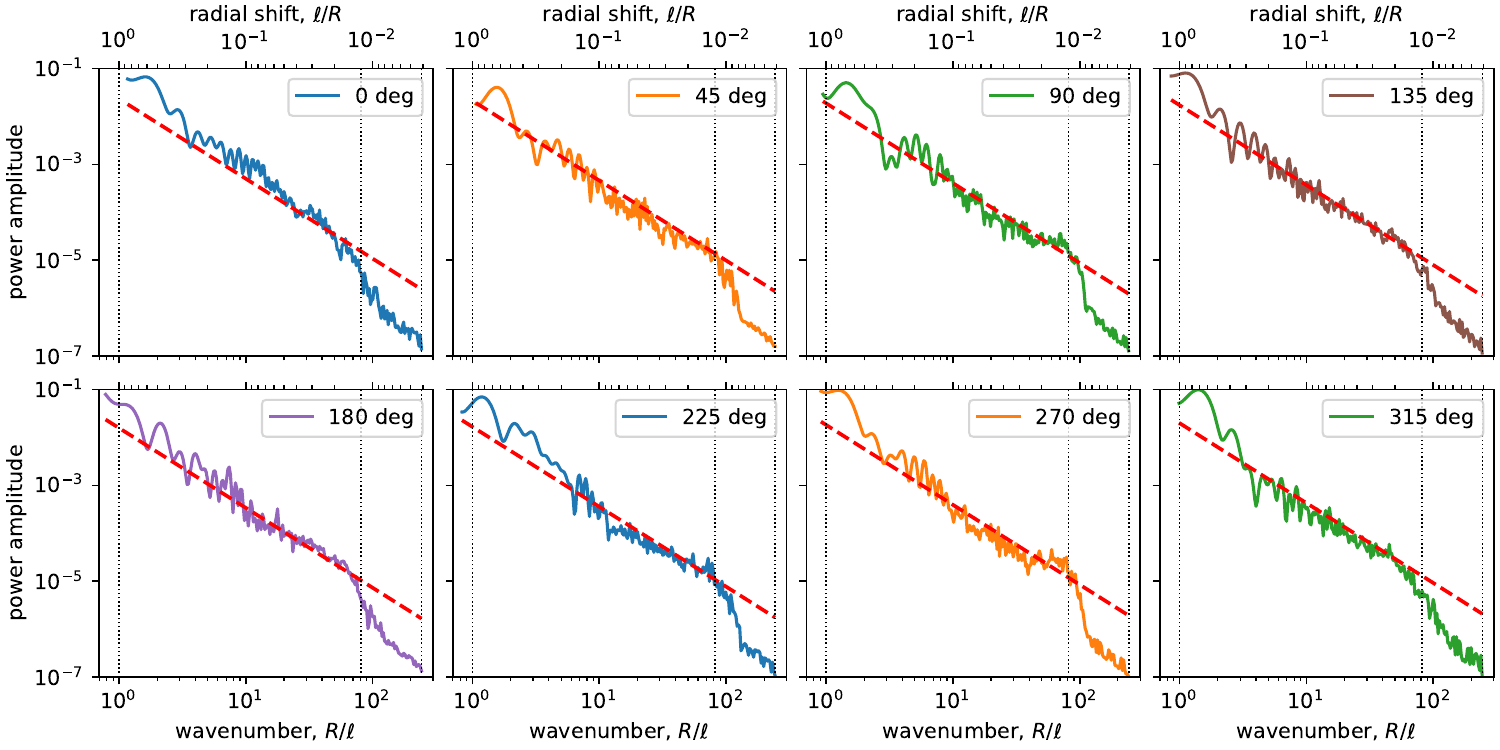}
  \caption{%
    Same as in Fig.~\ref{turb:ap:fig-sectors-xray-fft} but for the radio data.%
  }
  \label{turb:ap:fig-sectors-radio-fft}
\end{figure*}

Finally, Fig.~\ref{turb:ap:fig-sectors-xray-fft} and Fig.~\ref{turb:ap:fig-sectors-radio-fft} show the average power spectra for the same sectors. They are calculated as an average of the individual 45 brightness profiles in each sector. The continuous Fourier transform (through integration) was used. In contrast to the discrete Fourier transform, it accurately determines the location of modes at the non-integer wavenumbers in the spectrum. 




\bibliography{turbulence}{}
\bibliographystyle{aasjournal}

\end{document}